\documentclass[notoc,12pt]{article}
\pdfoutput=1
\usepackage{amsfonts}
\usepackage{amscd}
\usepackage{amssymb}
\usepackage{amsmath}
\usepackage{epsfig}
\usepackage{latexsym}
\usepackage{appendix}

\def \be  {\begin{equation}}
\def \ee  {\end{equation}}
\def \ba  {\begin{eqnarray}}
\def \ea  {\end{eqnarray}}
\def \bb  {\begin{thebibliography}}
\def \eb  {\end{thebibliography}}
\def \lab #1 {\label{#1}}

\newcommand\cA{\mathcal{A}}

\newcommand\C {\mathbb{C }}

\newcommand\rd{\mathrm{\, d}}

\newcommand\Li{\mathrm{Li}}

\newcommand\la{\langle}
\newcommand\ra{\rangle}

\newcommand\sgn{\mathrm{sgn}}

\newtheorem{lemma}{Lemma}

\begin{document}

\title{From $d\,$logs to dilogs; the super Yang-Mills MHV amplitude revisited}
\author{\normalsize Arthur E. Lipstein \& Lionel Mason \\ \small \textit{The Mathematical Institute} \\ \small \textit{University of Oxford} \\ \small \textit{24-29 St Giles'} \\ \small \textit{Oxford, OX1 3LB, U.K.}}
\maketitle
\begin{abstract}
Recently, loop integrands for certain Yang-Mills scattering amplitudes and correlation functions have been shown to be systematically  expressible in $d\,$log form, raising the possibility that these loop integrals can be performed directly without Feynman parameters.  We do so here to give a new description of the planar 1-loop MHV amplitude in $\mathcal{N}=4$ super Yang-Mills theory. We explicitly incorporate the standard Feynman $i\epsilon$ prescription into the integrands. We find that the generic MHV diagram contributing to the 1-loop MHV amplitude, known as Kermit, is dual conformal invariant up to the choice of reference twistor explicit in our axial gauge (the generic MHV diagram was already known to be finite).   The new formulae for the amplitude are nontrivially related to previous ones in the literature.  The divergent diagrams are evaluated using mass regularization. Our techniques extend directly to higher loop diagrams, and we illustrate this by sketching the evaluation of a non-trivial 2-loop example.  We expect this to lead to a simple and efficient method for computing amplitudes and correlation functions with less supersymmetry and without the assumption of planarity.      
{\normalsize \par}
\end{abstract}

\section{Introduction}

The planar 1-loop MHV amplitude of $\mathcal{N}=4$ super Yang-Mills (sYM) has proved to be one of the most important multipoint amplitudes.  It was first computed in \cite{Bern:1994zx} via unitarity methods as a sum of box functions of kinematic invariants, expressible as a sum of dilogs and infrared divergent terms.  It now provides a reference point for subsequent methods and techniques. Following Witten's discovery of twistor-string theory \cite{Witten:2003nn}, it was found that sYM amplitudes could be computed using the MHV diagram formalism (where Maximal Helicity Violating, or MHV, tree amplitudes are used as the Feynman vertices for constructing all other amplitudes \cite{Cachazo:2004kj,Risager:2005vk}).  The calculation of the planar 1-loop MHV amplitude based on the MHV diagram formalism in \cite{Brandhuber:2004yw} was the first clear indication that  this would work beyond tree level.  Subsequently, planar amplitudes and correlation functions of $\mathcal{N}=4$ sYM were found to have many more remarkable properties. In addition to superconformal symmetry, they also have dual superconformal symmetry \cite{Dualc,Drummond:2007aua,Brandhuber:2008pf,Drummond:2008vq} giving rise to Yangian symmetry \cite{Dolan:2004ps,Drummond:2009fd}. Dual superconformal symmetry stems from a duality between scattering amplitudes and null polygonal Wilson loops \cite{Alday:2007hr,Drummond:2007cf, Mason:2010yk,CaronHuot:2010ek,Bullimore:2011ni}; the dual superconformal symmetry for the amplitude is the ordinary superconformal symmetry for the Wilson-loop\footnote{There is furthermore a Wilson loop/correlator duality \cite{Alday:2010zy,Eden:2010zz,Eden:2010ce,Eden:2011yp,Eden:2011ku,Adamo:2011dq}.}.  A key early example of this duality was the computation of the 1-loop contribution to the null-polygonal Wilson loop \cite{Brandhuber:2007yx}, which gave the 1-loop MHV amplitude stripped of its  Parke-Taylor tree prefactor precisely.  

Further insights emerge when gauge theories are formulated in twistor space. In an axial gauge, the Feynman vertices of the twistor action correspond to MHV tree amplitudes and can be used to generate other amplitudes and correlation functions, providing the origin for the MHV diagram formalism   \cite{Mason:2005zm,Boels:2006ir,Boels:2007qn,Bullimore:2010pj,Mason:2010yk,Adamo:2011cb}. Furthermore, the momentum twistor space that makes dual conformal invariance manifest \cite{Hodges:2009hk} is the twistor space for the null polygonal Wilson loop dual to the scattering amplitudes. This Wilson loop can be re-expressed in momentum twistor space as a holomorphic Wilson-loop \cite{Mason:2010yk} that can also be computed by MHV diagrams.  The amplitude/Wilson loop duality is then simply realized as planar duality for the corresponding MHV-diagrams, which are the the twistor space Feynman diagrams in an axial gauge.  In approaches based on MHV diagrams, the axial gauge breaks dual conformal symmetry by virtue of a choice of reference twistor $Z_*$ that determines the direction in which the twistor space gauge field is trivialized. The loop integrands constructed from MHV diagrams are otherwise dual conformal invariant and manifest the appropriate permutation symmetries.  See \cite{Adamo:2011pv} for a review of these twistor related developments.

Recently it was discovered that for certain gauge theory scattering amplitudes and correlation functions, the loop integrands can be expressed in $d\,$log form.  More precisely, for MHV amplitudes, the loop integrand is a product of exterior derivatives of logarithms of rational functions and for higher MHV degree, the loop integrand consists of $d\, \log$ s multiplied by delta functions. Indeed, the Feynman rules for the twistor holomorphic Wilson loop can be rewritten in $d\, \log$  form \cite{Lipstein:2012vs}. These Feynman rules are very similar to those for the amplitude in twistor space as described in \cite{Adamo:2011cb} as they have the same origin in the twistor action. Furthermore, many of these ideas are are not restricted to planarity or maximal supersymmetry, so the $d\, $log form is likely to apply more generally.  

From the point of view of the twistor Wilson loop, the $d\,$log form of loop integrands of planar $\mathcal{N}=4$ sYM amplitudes  has a simple geometric interpretation:  the rational functions that are arguments of the $d\, $logs correspond to insertion points of propagators on MHV vertices or edges of the Wilson loop.  The external data of the scattering amplitude is encoded in the integration contours, which are given by reality conditions on the line determining the loop momentum in twistor space. An alternative $d\, \log$  form was obtained in \cite{ArkaniHamed:2012nw} from BCFW recursion using on-shell diagrams and the Grassmannan integral formula\footnote{ BCFW recursion relations (which relate higher point on-shell amplitudes to lower-point on-shell amplitudes \cite{Britto:2004ap,Britto:2005fq,ArkaniHamed:2010kv})  can also be simply realised in twistor space \cite{Mason:2009sa,ArkaniHamed:2009si}, expressing full superconformal invariance.  This  led to a formula that generates the amplitudes and leading singularities of $\mathcal{N}=4$ sYM via a contour integral over a Grassmannian \cite{ArkaniHamed:2009dn}.  There is a parallel Grassmannian dual conformal invariant formula for the Wilson loop \cite{Mason:2009qx}, and the translation between the two expresses the Yangian symmetry  in this framework \cite{ArkaniHamed:2009vw}.  BCFW recursion was extended to generate the loop integrand in \cite{ArkaniHamed:2010kv,Boels:2010nw}.    }. In the alternative approach, the rational functions in the $d\,$log form correspond to BCFW shifts which arise from an on-shell diagram formulation of  the all-loop recursion relations.  In both approaches, it is clear that the $d\,$log form of the loop integrand is not really tied to planarity, nor to maximal supersymmetry, but in this paper we nevertheless focus on these most basic and simplest examples.

The loop integrand is not however the main objective; one is really after fully integrated amplitudes and correlation functions.  The attraction of the $d\, $log form of the integrand is that it is very suggestive for direct integration without Feynman parameters. In this form, the integrand is locally exact so it should be possible to iteratively use Stokes theorem to reduce integration to combinatorics. The integrals are still nontrivial to perform, however, since one must take into account the contour of integration. Indeed, this is where the kinematic data is encoded. Furthermore, the singularities made explicit in the $d\,$log form mean that the answer very much depends on the contour. In our formulation, the integrand of an $L$-loop MHV amplitude consists of $d\,$log's of $4L$ variables, and half of these are constrained to be real by the reality constraints in momentum twistor space.  For a complex variable $s$, $d\, \log s=d s/s$ is integrable and well-defined, but for $s$ real, the integral of $d \log s$  is not well-defined unless regulated by some $i\epsilon$ prescription. We will see that these real integration variables are directly related to physical propagators in space-time and so we must use the Feynman $i\epsilon$ prescription to make these real integrals well defined.  Another puzzle is that the degree of transcendentality for an $L$-loop amplitude is expected to be $2L$, whereas naively the integral of an expression in $4L$ $d\, $logs would usually define a polylogarithm expression of transcendentality degree $4L$.  The resolution of this is that the usual definition of $Li_k$, which has of transcentality degree $k$, involves $k$ iterated {\em indefinite} integrals, but our integral is compact with no boundary.  We will see that in the integration procedure, half the $d\,$logs are used to restrict the contour to one that is a product of $2L$ intervals with the remaining integrand in $d\,$log form as required for transcendality degree $2L$.

Here we develop a systematic method for evaluating the integrals of loop amplitudes from the $d\,$log form of their integrands.  We focus on the 1-loop MHV amplitude, for which we give a complete treatment. In this case there is only one contributing MHV-diagram $K_{ij}$, where $i,j=1,\ldots,n$ index the external legs of the amplitude,  that has become known as Kermit\footnote{because of the diagram's resemblance to a puppet frog of the same name.}.  The integrand for $K_{ij}$ is the same as the integrand for the 1-loop MHV amplitude obtained by \cite{ArkaniHamed:2010kv} via BCFW recursion, although this equivalence between MHV-diagrams and BCFW expressions is a coincidence that does not persist to higher loop orders or MHV degree. 
The full planar 1-loop MHV amplitude is given by
\be\label{MHV0}
\cA(1,\ldots,n)=\frac12 \sum_{i,j}K_{ij}.
\ee
  In the generic case  $ |i-j|\geq 2 $, 
our new formulae for $K_{ij}$ are
\begin{equation}
K_{ij}=\Li_{2}\left(\frac{a_{i\, j}}{v_{*}}\right)+\Li_{2}\left(\frac{a_{i-1\, j-1}}{v_{*}}\right)-\Li_{2}\left(\frac{a_{i-1\, j}}{v_{*}}\right)-\Li_{2}\left(\frac{a_{i\, j-1}}{v_{*}}\right)
 + c.c.\ 
\label{kermitgeneric}
\end{equation}
where $a_{ij}=\la i|x_{ij}\left|j\right]/(\left\langle i \bar{\eta}\right\rangle \left[\eta j\right])=i Z_i\cdot\bar Z_j $, $v_*= x_{ij}^2/2\la \bar{\eta}|x_{ij}|\eta]$, and $\eta$ is the reference spinor. We have normalized the external twistors so that $\la i \bar{\eta}\ra=Z_i\cdot \bar Z_*=1$. If we use \eqref{eq:vstar} for $v_*$, we see that this has manifest (dual) conformal symmetry up to the choice of a reference twistor $Z_*$ (which encodes the reference spinor). Dual conformal invariance is broken only in the divergent diagrams when $|i-j| = 1$.   We regulate this case by taking $i \epsilon \rightarrow i \epsilon - m^2$ when implementing the $i \epsilon$ prescription, where $m^2$ is the mass parameter  for a mass regulation. We find that
\begin{equation}
K_{ii+1}=-\frac{1}{4}\left(\ln^{2}\left(\frac{m^{2}}{x_{ii+2}^{2}}\right)+\ln\left(\frac{x_{i-1i+1}^{2}}{x_{ii+2}^{2}}\right)\ln\left(x_{ii+2}^{2}\right)\right)-\frac{2\pi^2}{3}+\mathcal{O}(m). 
\label{kermitdivergent}
\end{equation}
Finally, when $i=j$, Kermit vanishes $K_{ii}=0$. This result has the standard divergent behaviour. 

Our results for the planar 1-loop MHV amplitude of $\mathcal{N}=4$
 sYM are nontrivially related to previous formulae. The formulae
 obtained in \cite{Bern:1994zx} expressed the answer in terms 
of a sum of two mass easy box functions, each a sum of 5 dilogs. In \cite{Brandhuber:2004yw,Brandhuber:2007yx}  the two-mass easy box function was reduced to a simpler form containing just four dilogarithms via a 9 term dilog identity.  Although the analysis in \cite{Brandhuber:2004yw} verified the applicability of the MHV-diagram formalism for these loop amplitudes, it was not possible in that analysis to derive an expression for a single Kermit diagram since different choices of reference spinor were made for different cuts contributing to a single Kermit. Indeed, the results in \cite{Brandhuber:2004yw} do not have manifest dual conformal invariance and non-trivial dilog identities must be used to relate our formulae to theirs as we see in \ref{BSTa}. In appendix \ref{massdetails}, we verify that \eqref{kermitdivergent} reproduces the form of the 1-loop 4-point MHV amplitude obtained in \cite{Alday:2009zm}.

The method for obtaining the 1-loop MHV amplitude described here is substantially simpler than previous methods. We expect that the techniques developed in this paper will extend to higher-loop scattering amplitudes and correlation functions in $\mathcal{N}=4$ sYM as well as to other field theories.  As far as the higher-loop MHV amplitude is concerned, our argument in \cite{Lipstein:2012vs} for reduction of an $L$-loop MHV integrand into $d\, $log form first showed that it could be built essentially from $L$ Kermits, albeit with some as arguments of others.  We illustrate this  in section \ref{higherloops}, where we sketch the evaluation of a nontrivial 2-loop diagram.

This paper is organized as follows. In section \ref{background}, we review momentum twistors and the $d\, \log$  form of the planar 1-loop MHV amplitude of $\mathcal{N}=4$ sYM. In section \ref{generic1}, we we explain how to implement the Feynman $i \epsilon$ prescription and present the result for generic generic Wilson loop diagrams contributing to the 1-loop MHV amplitude. In section \ref{divergent}, we describe how to regulate divergent Wilson loop diagrams using mass-regularization. In section \ref{higherloops}, we explain how our methods can be applied to higher loop planar amplitudes, and illustrate this by giving a preliminary discussion of a nontrivial two-loop example. In section \ref{conclusion}, we present some conclusions and describe some future directions. In appendix \ref{BSTa}, we show that our result for the generic Wilson-loop diagram is equivalent to the result for the 1-loop MHV amplitude previously obtained in \cite{Brandhuber:2004yw} using unitarity methods applied to the MHV diagram formalism. In appendix \ref{massdetails} we provide more details about mass-regularization of divergent Wilson loop diagrams. In appendix \ref{symbol}, we reduce the symbol of the 1-loop MHV amplitude to a sum over terms that consist of the ingredients in Kermit as used in \S \ref{generic1} to show that our generic term gives rise to the correct symbol up to terms that cancel telescopically.                 

\section{Background} \label{background}

In this section, we review the basic definitions and set up the notation for the rest of the paper. First we review variables which are useful for computing planar on-shell scattering amplitudes in $\mathcal{N}=4$ sYM, notably region momenta and momentum twistors. Then we review the 1-loop MHV amplitude in $\mathcal{N}=4$ sYM, in particular the $d\, \log$  form of its loop integrand.       

\subsection{Momentum Twistors}

We first write on-shell momenta in two-component spinor form as
follows: \[
p^{\alpha\dot{\alpha}}=\lambda^{\alpha}\tilde{\lambda}^{\dot{\alpha}}\]
where $\alpha=0,1$ and $\dot{\alpha}=\dot{0},\dot{1}$ are chiral
and antichiral spinor indices. The particles
also have fermionic supermomentum $$q^{\alpha a}=\lambda^\alpha \rho^a,$$ where $a$ is an $SU(4)$ R-symmetry
index and $\rho$ is fermionic. 
The $n$-point superamplitudes are then functions of $n$ such $(\lambda,\rho)$ variables, one for each external particle. For example, a tree-level $n$-point MHV superamplitude has the following simple form 
\begin{equation}
A_{n}^{MHV}=\frac{\delta^{4}\left(P\right)\delta^{8}\left(Q\right)}{\left\langle 12\right\rangle \left\langle 23\right\rangle ...\left\langle n1\right\rangle }
\label{MHV}
\end{equation}
where $P=\sum_{i=1}^{n}p_{i}$, $Q=\sum_{i=1}^{n}q_{i}$, and $\left\langle ij\right\rangle =\epsilon_{\alpha\beta}\lambda_{i}^{\alpha}\lambda_{j}^{\beta}$.

Dual superconformal symmetry for planar $\mathcal{N}=4$ sYM amplitudes can be seen by
arranging the external supermomenta of a colour ordered amplitude into a polygon
and writing the amplitude as a function of the vertices of this polygon, which lives in `region momentum space'. Dual superconformal symmetry then corresponds to ordinary conformal symmetry
in region momentum space. The region momentum space coordinates 
are defined by 
\begin{equation}
\left(x_{i}-x_{i+1}\right)^{\dot{\alpha}\alpha}=\lambda_{i}^{\alpha}\tilde{\lambda}_{i}^{\dot{\alpha}},\,\,\,\left(\theta_{i}-\theta_{i+1}\right)^{a\alpha}=\lambda_{i}^{\alpha}\rho_{i}^{a}.\label{eq:dual}\end{equation}
This representation automatically incorporates momentum conservation. The duality between amplitudes and Wilson-loops is the equivalence of the planar scattering amplitude with the planar Wilson-loop around this polygon.  

The dual superconformal symmetry of the amplitudes can be made more manifest
by writing the polygon in terms of (momentum) supertwistors:
\[
\left(Z_{i}^{A},\chi_{i}^{a}\right)=\left(\lambda_{i\alpha},\mu_{i}^{\dot{\alpha}},\chi_{i}^{a}\right).\]
They transform in the fundamental representation of the dual superconformal group $SU(2,2|4)$ and relate to the region supermomenta by the
`incidence relations' \begin{equation}
\mu_{i}^{\dot{\alpha}}=-ix_{i}^{\dot{\alpha}\alpha}\lambda_{i\alpha},\,\,\,\chi_{i}^{a}=-i\theta_{i}^{a\alpha}\lambda_{i\alpha}.\label{eq:incidence}\end{equation}
These express a point $(x,\theta)$ in space-time as a complex projective line $X$ in twistor space, i.e. the point  $x_i$ corresponds to the line $X_i$ passing through both $Z_i$ and $Z_{i-1}$.  
Such a projective line in
twistor space can be represented as a skew twistor
\begin{equation}
X_{i}^{AB}=\frac{Z_{i}^{[A}Z_{i-1}^{B]}}{\left\langle ii-1\right\rangle }
\label{bigX1}
\end{equation}
where we have normalized using the spinor brackets in the denominator.  Although the skew twistor is conformally invariant up to scale, its normalisation is not and requires the knowledge of the `infinity twistor' $I_{AB}$ defined by
\[
\left\langle ij\right\rangle =I_{AB}Z_{i}^{A}Z_{j}^{B},\,\,\, I_{AB}=\left(\begin{array}{cc}
\epsilon^{\alpha\beta} & 0\\
0 & 0\end{array}\right).\]
Given this, setting $x_{ij}=x_i-x_j$, the distance between two points in the dual space can be written in
terms of momentum twistors as follows: 
\[
x_{ij}^2:=\left(x_{i}-x_{j}\right)^{2}=\frac{\left( ii-1jj-1\right) }{\left\langle ii-1\right\rangle \left\langle jj-1\right\rangle },\,\,\,\left( ijkl\right) =\epsilon_{ABCD}Z_i^{A}Z_j^{B}Z_k^{C}Z_l^{D}.\]

If the spacetime has Lorentzian signature, the complex conjugate of
a twistor $Z^{A}$ is a dual twistor given by $\bar{Z}_{A}=\left(\bar{\mu}^{\alpha},\bar{\lambda}_{\dot{\alpha}}\right)$.  The reality condition that the polygon lies in real Minkowski space is that
$$
\bar Z_{iA}=\alpha\varepsilon_{ABCD}Z_{i-1}^BZ_i^CZ_{i+1}^D\, ,
$$
for some $\alpha\neq 0$.

In the MHV diagram formalism, we also have a reference spinor $\eta^{\dot \alpha}$ which, in momentum twistors we incorporate as a reference twistor
$$
Z_*=(0,\eta^{\dot\alpha})\, .
$$
Ordinarily the twistors $Z_i$ do not have a natural choice of scaling (just as $\lambda_i$ does not).  However, we will find it convenient in what follows to normalize all the $\lambda_i$ and hence $Z_i$ so that 
$$
\la \lambda_i\,\bar \eta\ra=Z_i\cdot \bar Z_*=1\, .
$$
With this, we can now form the invariants 
\begin{equation}
a_{ij}:=iZ_i\cdot\bar Z_j =\frac{\left\langle i\right|x_{ij}\left|j\right ] }{\left\langle i \bar{\eta}\right\rangle \left[\eta j\right]},
\label{aijs}
\end{equation}
where we have included the factors in the denominator in order to indicate our normalization of the external twistors. For simplicity, we will set these normalization factors to 1 in the remainder of the paper. These invariants are, up to normalisation, the same as the invariants $(i\, j-1\, j\, j+1)$ used elsewhere, e.g.\ \cite{CaronHuot:2011ky,Golden:2013lha}. Note that they differ by a factor of $i$ from those used in \cite{Lipstein:2012vs}. These depend also on the reference twistor, and only those combinations that are independent of the choice of the scalings of the $Z_i$ are fully (dual-) conformal invariant. However, we cannot expect to obtain quantities that are independent of the  reference twistor from MHV diagrams in the first instance, so the $a_{ij}$ are a natural set of kinematic variables for us.  

\subsection{The 1-loop MHV Amplitude}

\begin{figure}[h]
\begin{center}
\includegraphics[scale=0.18]{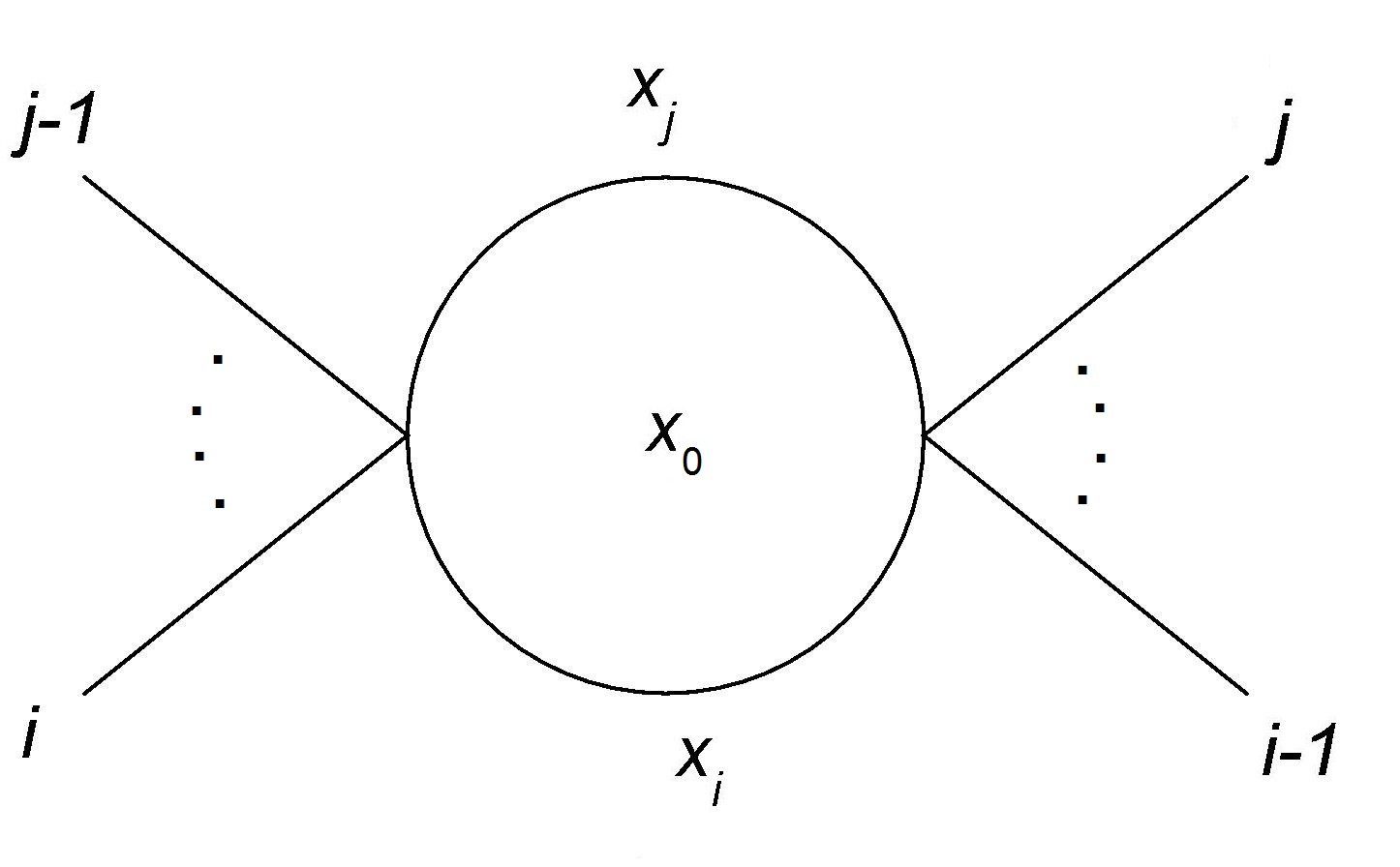}
\caption{1-loop MHV diagram in the CSW formalism.}
\label{1lmhv}
\end{center}
\end{figure}

In the MHV diagram formalism, loop amplitudes in $\mathcal{N}=4$ sYM can be computed using the tree-level MHV amplitudes \eqref{MHV} as the Feynman vertices with scalar propagators. The diagrams for the planar 1-loop MHV amplitude are obtained by connecting two MHV vertices by two propagators, and a generic example is depicted in Figure \ref{1lmhv}. The external region momenta  as defined by \eqref{eq:dual} are supplemented by $x_0$ for the loop momentum and the regions $(x_i,x_j)$, which are adjacent to the two propagators. The full planar 1-loop MHV amplitude is given by summing over all $(i,j)$ with $i\neq j$. As for an amplitude, a vertex must have at least three legs.  The MHV vertices  \eqref{MHV} are extended off-shell so that the off-shell momentum $x_{0i}$ corresponds to the spinor $[\eta |x_{0i}$ and so on. 
   
\begin{figure}[h]
\begin{center}
\includegraphics[scale=0.8]{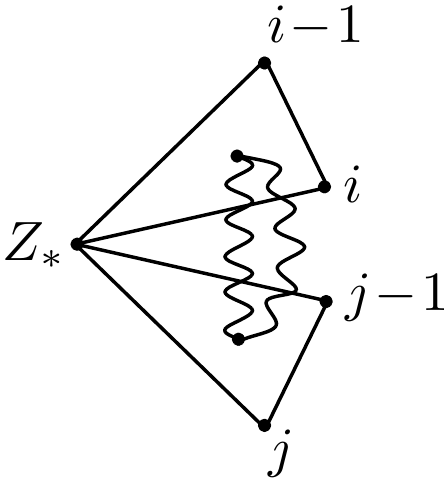}
\caption{Kermit diagram for 1-loop MHV diagram.}
\label{1lmhv-kermit}
\end{center}
\end{figure}

After stripping off the MHV tree amplitude, the diagram in Figure \ref{1lmhv} is given by \cite{Brandhuber:2004yw} 
\begin{equation}
K_{ij}=\int  \frac {\la i-1\, i\ra \la j-1\, j\ra \rd^4 x_0}{x_{0i}^2 x_{0j}^2 \la i-1|x_{0i}|\eta] \la i|x_{0i}|\eta] \la j-1|x_{0j}|\eta]\la j|x_{0j}|\eta]}
\label{kermitr}
\end{equation}
where $\eta$ is the reference spinor and $x_{ij}=x_i - x_j$. The object $K_{ij}$ is sometimes referred to as Kermit. In terms of momentum twistors, Kermit is given by \cite{Bullimore:2010pj} 
\begin{equation}
\int \frac{\rd^{4|4} Z_A\rd^{4|4}Z_B  \left(\left(*i-1i\left[A\right.\right)\left(\left.B\right]j-1j*\right)\right)^{2}}{(\mbox{Vol GL}_2 )(A\, B \, i-1\, i)(A\, B\, j-1\, j)(A\, B\, *\, i-1)(A\, B\, *\, i)(A\, B\, *\, j-1)(A\, B\, *\, j)}
\label{kermittwist}
\end{equation}
where the reference spinor $\eta$ has been embeded in the reference twistor 
$
Z_*=(0,\eta^{\dot \alpha})\, ,
$
 the loop region momentum $x_0$ corresponds to the line spanned by the twistors $(Z_A,Z_B)$, and the $GL_2$ is associated with the choice of $(Z_A,Z_B)$ from within their span and is understood to be fixed by a standard Fadeev-Popov procedure.  The singularity structure is given by the intersection of the line $X_0$ with the solid lines in twistor space depicted in figure \ref{1lmhv-kermit}  (the wiggly lines correspond to the numerator factors) which has a superficial similarity to Kermit the frog. The integrand in \eqref{kermittwist} is the same as the integrand for the 1-loop MHV amplitude obtained using all-loop BCFW recursion in \cite{ArkaniHamed:2010kv}. In that reference, the role of the reference twistor is played by $Z_1$.

The $d\, $log form of the integrand for Kermit as given in \cite{Lipstein:2012vs} is most easily expressed in terms of the holomorphic Wilson loop in twistor space that is dual to the planar S-matrix of $\mathcal{N}=4$ sYM. In particular, the arguments of the $d\, $logs are the coordinates corresponding to insertion points of  propagators onto MHV vertices or edges of the Wilson loop. We now briefly review how this arises for Kermit, see \cite{Lipstein:2012vs} for further details.   
\begin{figure}[h]
\begin{center}
\includegraphics[scale=0.14]{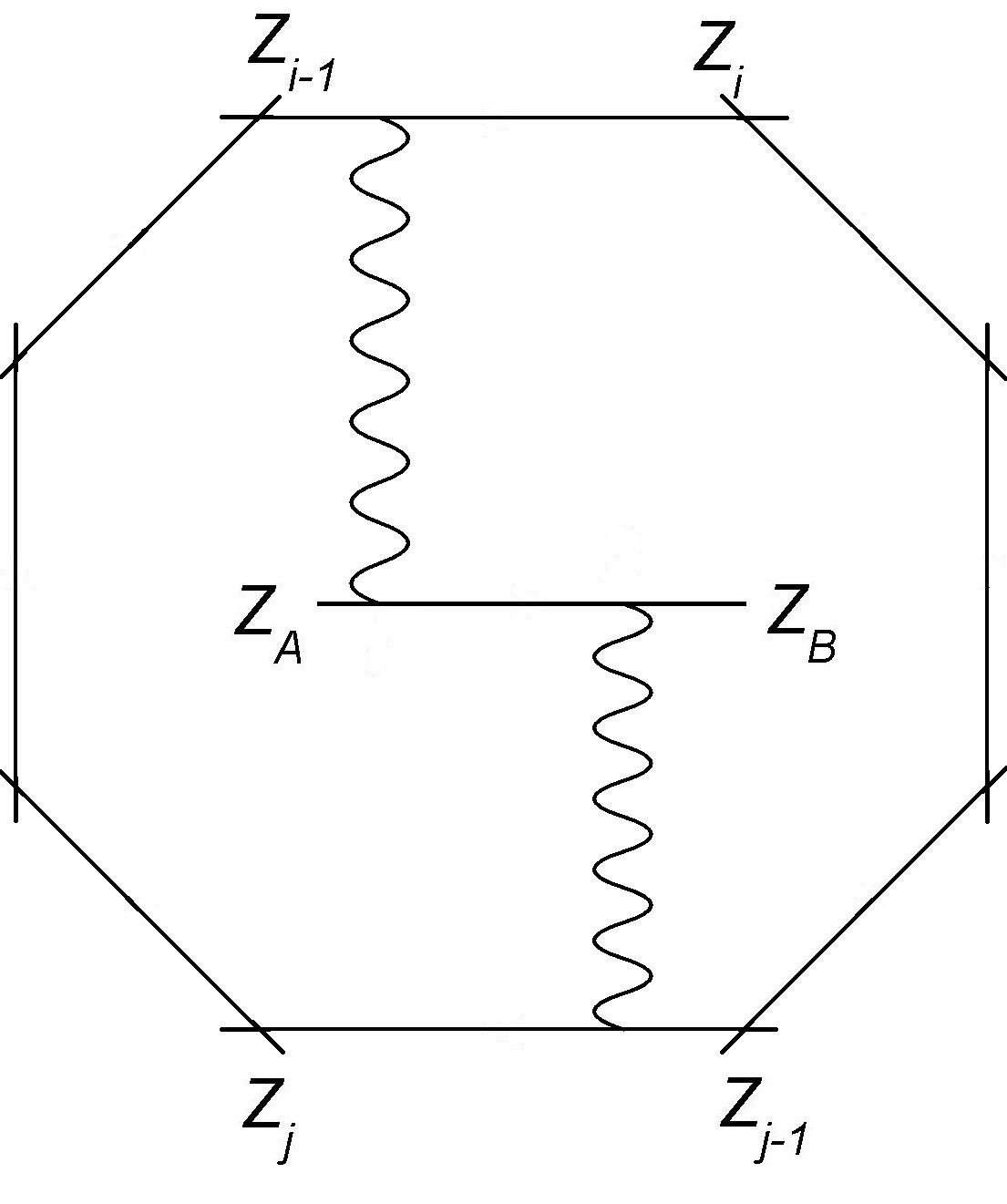}
\caption{A generic Wilson loop diagram contributing to the 1-loop MHV amplitude.}
\label{fig1loop}
\end{center}
\end{figure}
The twistor Wilson loop Feynman diagram that corresponds to Kermit is given in Figure \ref{fig1loop}. It is related to Figure \ref{1lmhv} by planar duality, which exchanges loops in the amplitude diagram with MHV vertices in the Wilson loop diagram, as depicted in Figure \ref{duality}. Hence the loop region momentum $x_0$ in Figure \ref{1lmhv} corresponds to the line MHV vertex in Figure \ref{fig1loop} supported on the line in twistor space spanned by the twistors $(Z_A,Z_B)$.  The wavy lines in Figure \ref{fig1loop} correspond to propagators in twistor space\footnote{More generally, a twistor Wilson loop diagram with $L$ MHV vertices will correspond to an $L$-loop amplitude. In general, a twistor Wilson loop diagram with $P$ propagators and $L$ MHV vertices, corresponds to an $L$-loop $N^k$MHV amplitude, where $k=P-2L$.}.      
\begin{figure}[h]
\begin{center}
\includegraphics[scale=0.8]{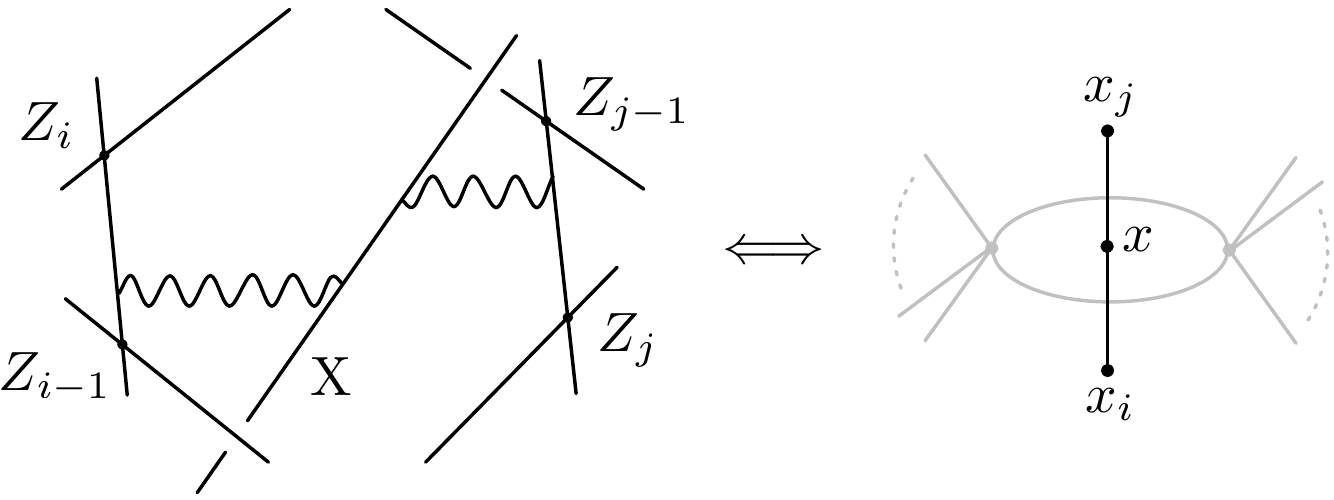}
\caption{Amplitude/Wilson-loop duality for 1-loop MHV diagram.}
\label{duality}
\end{center}
\end{figure}

Propagators in the twistor Wilson loop diagram correspond to delta functions
$$
\Delta(Z,Z'):=\frac1{2\pi i}\bar\delta^{2|4}(Z,Z_*,Z'):=\frac1{2\pi i}\int_{\C^2} \frac{\rd u}{u} \frac{\rd v}{v}\bar\delta^{4|4}(Z+u Z_* + vZ')
$$ 
in twistor space which is essentially a delta function restricting $Z$ and $Z'$ to lie along a line through $Z_*$ and a Cauchy pole along that line when $Z$ meets $Z'$. As explained in \cite{Lipstein:2012vs}, it is possible to define the parameters in the propagators and fix the $GL_2$ freedom so that on integrating out the $Z_A$ and $Z_B$, the delta functions enforce  $Z_A$ and $Z_B$ to be
\be
Z_A=is_0Z_* +  \frac {Z_{i-1} +s Z_{i}}{1+s} \, , \qquad Z_B=it_0 Z_* +  \frac{ Z_{j-1} +t Z_{j}}{1+t}\, 
\label{zazb}
\ee
where $(s_0,t_0,s,t)$ are rational functions of the loop momentum that can be used as alternative integration variables for the twistor Wilson loop diagram. In terms of $(s_0,t_0,s,t)$, the Kermit integral reduces to 
\be{}
K_{ij}=-\frac{1}{4\pi ^2}\int
 \rd \ln{s_0}\,\rd \ln{t_0} \, \rd \ln{s} \,\rd \ln{t}\, .
\label{dlog}
\ee
Now the information of the external twistors is encoded into the
contour, i.e., the condition that the line joining $Z_A$ to $Z_B$ is real.  
We  normalize the external twistors so that $$Z_i\cdot \bar Z_*=1\, .$$
With this, the reality conditions are 
\be\label{real0}
0= Z_A\cdot \bar Z_A= i(s_0-\bar s_0)\, ,  \qquad 0=Z_B\cdot \bar Z_B=i(t_0-\bar t_0)\, , 
\ee
and 
\be\label{real1}
 0=Z_A\cdot \bar Z_B= i(s_0 - \bar t_0) -i \frac{  a_{i-1\, j-1} + s
   a_{i\, j-1} +\bar t a_{i-1\, j} + s\bar t a_{ij}}{(1+s)(1+\bar t)}\, ,
\ee
where we have set $a_{ij}=iZ_i\cdot\bar Z_j$. A more explicit formula for these invariants is given in \eqref{aijs}.
Thus the reality conditions imply that $s_0$ and $t_0$ are real and express $s$ as a Mobius transform of $\bar t$, which depends on $v= s_0-t_0$:
\begin{equation}
s=-\frac{\bar{t}\left(a_{i-1j}-v\right)+a_{i-1j-1}-v}{\bar{t}\left(a_{i\, j}-v\right)+a_{ij-1}-v}\, .
\label{eq:real}\end{equation}

The $(s_0,t_0,s,t)$ can be expressed as functions of $(x_0, x_i, x_j)$ as follows:
\be\label{coord-def}
s_0 =\frac{- x_{0i}^2}{2[\eta|x_{0i}|\bar\eta\ra}\, , \quad t_0= \frac{- x_{0j}^2}{2[\eta|x_{0j}|\bar\eta\ra}\, , \quad s= \frac {\la i-1|x_{0i}|\eta]}{\la i|x_{0i}|\eta]}\, ,\quad t= \frac {\la j-1|x_{0j}|\eta]}{\la j|x_{0j}|\eta]}\, ,
\ee
and these can be expressed dual conformal invariantly as 
\be
s_0 =\frac{(Z_AZ_B  Z_{i-1}Z_i)}{(Z_* Z_A Z_B  X_i \cdot \bar Z_*)}\, , \quad s= \frac {(Z_{i-1} Z_AZ_B  Z_*)}{(Z_ i Z_AZ_BZ_*)}\, , \quad \mbox{ and } \quad i, s \leftrightarrow j, t \, .
\ee
These follow from the incidence relations, which correspond to
\begin{eqnarray}\nonumber
Z_A&=&(\lambda_A,-i x_0|\lambda_A\ra), \qquad Z_B=(\lambda_B,-ix_0 |\lambda_B\ra)\, , 
\\ Z_{i-1}&=&(\lambda_{i-1},-ix_i|i-1\ra)\, , \quad Z_i=(\lambda_i, -ix_i |i\ra)\, ,
\end{eqnarray}
and $i\leftrightarrow j$, together with $Z_*=(0,\eta)$ and $\la \lambda_i\, \bar\eta\ra=1$.  Writing \eqref{zazb} in  spinor parts gives
\be\label{incidence}
\lambda_A= \frac{\lambda_{i-1}+s\lambda_i}{1+s} \, \quad \lambda_B=\frac{\lambda_{j-1}+t\lambda_j}{1+t} \, , \quad x_{0i}|\lambda_A\ra =-s_0 \eta\, , \quad x_{0j}|\lambda_B\ra=-t_0\eta.
\ee
Multiplying the last two equations by $x_{0i}$ and $x_{0j}$ respectively and taking various components etc., leads to \eqref{coord-def}.

The $d\, \log$  form of the loop integrand in \eqref{dlog} is  a coordinate transformation of the original Kermit integrand in \eqref{kermitr}.
However, clearly $(s_0,t_0,s,t)$ are redundant coordinates for the real slice, so we will eliminate $s$ in favour of $t$ using the reality condition in \eqref{eq:real}.  We can then express $x_{0j}$ in terms of $(s_0,t_0,t)$ directly   by
\be\label{t-coords}
x_{0j}= \frac { x_{ij}^2/2-v x_{ij}\cdot n }{[\bar \lambda_B|x_{ij}|\lambda_B\ra -v} \bar \lambda_B\lambda_B + t_0 n\, , \quad \mbox{ where } \quad n=\bar \eta \eta\, , \quad v=s_0-t_0.
\ee
where $\lambda_B$ is given in terms of $t$, $\lambda_{j-1}$ and $\lambda_j$ in \eqref{incidence} above.  This is obtained by solving  the second and fourth equation in \eqref{incidence}, and the difference of the first two equations of \eqref{coord-def} for $x_{0j}$. Clearly a similar formula can be obtained in terms of $s$.

\section{Evaluating Kermit} \label{generic1}

The $d\, $log form  of the Kermit integrand is given in \eqref{dlog} with  contour \eqref{real0} and \eqref{eq:real}. The variables $s$ and $t$ are complex and so $\rd s/s$ and $\rd t/t$ are integrable and unambiguous.  However, $(s_0,t_0)$ are real so that the poles at $s_0,t_0=0,\infty$  must be regulated. From \eqref{coord-def}, we see that the $(s_0,t_0)$ poles are related to those of physical propagators in space-time. Hence, the Feynman $i\epsilon$ prescription for these poles will shift $(s_0,t_0)$ into the complex plane. After doing so, these real integrals will become well-defined.

\subsection{On the $i\epsilon$ prescription and the real integrals}

In terms of region momenta, the Feynman $i\epsilon$ prescription is simply
\be{}
\frac 1{x_{0i}^2}\rightarrow \frac 1{x_{0i}^2+ i \epsilon }\, , \qquad \frac 1{x_{0j}^2}\rightarrow \frac 1{x_{0j}^2+ i \epsilon }\, .
\ee
This will follow from \eqref{coord-def} if we take
\be{}
s_0 \rightarrow s_0 + i\epsilon f_i , \quad t_0\rightarrow t_0 + i \epsilon  f_j
\qquad \mbox{ where } \qquad 
f_i = \frac{-1}{2x_{0i}\cdot n} \, , \quad f_j=\frac {-1}{2x_{0j}\cdot n }\, .
\label{iepshift}
\ee
In terms of the $(v,t)$ coordinates we obtain
\be \label{fifj}
f_i=\frac {v-[\bar \lambda_B|x_{ij}|\lambda_B\ra}{x_{ij}^2-2x_{ij}\cdot n [\bar\lambda_B|x_{ij}|\lambda_B\ra } \, \qquad f_j= \frac {v-[\bar \lambda_B|x_{ij}|\lambda_B\ra}{x_{ij}^2-2v x_{ij}\cdot n }.
\ee

A key point is that there is some decoupling in the integral because, from \eqref{t-coords} and \eqref{iepshift}, $f_i$ and $f_j$ depend only on $v=s_0-t_0$ and not $s_0+t_0$, so that  we can perform the $t_0$ integral while holding $v$ and $t$ constant.  Setting $s_0=v+t_0$ we are therefore left with an integral of the form
\be
\int \frac {\rd t_0 \rd v}{(t_0+i\epsilon f_j) (v+t_0+ i\epsilon f_i)} F(v,t,\bar t)\rd t \rd \bar t\, ,
\ee
where $F$, $f_i$ and $f_j$, do not depend on $t_0$.  
If $f_i$ and $f_j$ have the same sign, then the contour can be contracted in the upper or lower half plane and the answer vanishes.  If they have opposite signs the contour integral picks up the residue at $- v$ or $0$ accordingly. The answer therefore reduces to
\be
2\pi i \int_{-\infty}^\infty \frac{\sgn f_j \, \theta(-f_if_j) \, \rd v}{v + i\epsilon(f_i-f_j)} F(v,\ldots)\, ,
\label{uvint0}
\ee
where $\theta$ is the Heavyside step function.

In coordinates in which $s$ is eliminated in favour of $t$,  $f_i$ and $f_j$ depend on $v$ and the step function  provide limits for the $v$ integration as follows.  Firstly, one can see immediately from their definitions that $f_i$ and $f_j$ are positive to the future of the hypersurfaces at $x_{0i}\cdot n=0$ and $x_{0j}\cdot n=0$ respectively and negative to their past.  Thus $\theta(-f_if_j)$ is supported in-between the two hypersurfaces and does not include infinity.  Thus, $f_j$ has a constant sign on the support of $\theta(-f_i f_j)$ and indeed its sign is that of $x_{ji}\cdot n$. This sign, being constant, can be taken out of the integrand.  
It follows from \eqref{t-coords} that the hypersurfaces $x_{0i}\cdot n=0$ and $x_{0j}\cdot n=0$ give the following  limits on the $v$-integral
\be
x_{0j}\cdot n= \frac {-v x_{ij}\cdot n + x_{ij}^2/2}{[\bar\lambda_B|x_{ij}|\lambda_B\ra-v}=0  \quad \mbox{ when } \quad v=v_*:= \frac{x_{ij}^2}{2x_{ij}\cdot n}
\ee  
and
 \be
 x_{0i}\cdot n= x_{0j}\cdot n -x_{ij}\cdot n=0 \quad \mbox{ when } \quad v=\pm \infty\, .
 \ee
 There is the possibility of a contribution from the pole at $v=0$.  However at $v=0$
 \be
 x_{0i}\cdot n\, x_{0j}\cdot n= \frac{x_{ij}^2(x_{ij}^2-2x_{ij}\cdot n [\bar\lambda_B|x_{ij}|\lambda_B\ra)}{4 [\bar\lambda_B|x_{ij}|\lambda_B\ra^2} =-x_{ij}^2  \frac{|[\eta|x_{ij}|\lambda_B\ra|^2 }{2 [\bar\lambda_B|x_{ij}|\lambda_B\ra^2}
\ee
 and this is positive in the Euclidean kinematic region usually considered, i.e., $x_{ij}^2<0$.  Thus $v=0$ lies outside the integration region and so does not contribute.  
 
 We will see later that the integrand vanishes as $v\rightarrow \infty$, so no more prescriptions to regulate the real integral need to be made and we can now set $\epsilon=0$ (except for the divergent diagrams that will be considered separately).  Finally, noting that $\sgn \, v_*= \sgn \, x_{ij}\cdot n$ in our kinematic region, we see that   the 
 real integral reduces to
\be
 2\pi i \int_{v_*}^{ \infty }\frac{\, \rd v}{v} F(v,\ldots)\, ,\quad v_*>0 \qquad \mbox{ or } \quad    -2\pi i\int_{-\infty}^{ v_* }\frac{\, \rd v}{v} F(v,\ldots)\, , \quad v_*<0\, .
\label{uvint}
\ee
We will take the former case for definiteness in what follows.

As a final remark in this subsection we note that critical value $v_*$ of $v$ can be expressed dual conformal invariantly, up to the choice of the reference twistor as
\begin{equation}
v_{*}=\frac{X_{i}\cdot X_{j}}{\bar{Z}_{*}\cdot X_{i}\cdot X_{j}\cdot Z_{*}}=\frac{a_{ij}a_{i-1j-1}-a_{ij-1}a_{i-1j}}{a_{ij} + a_{i-1j-1}-a_{i-1j}-a_{ij-1}}\, .\label{eq:vstar}\end{equation}
Thus we have implemented the standard Feynman $i\epsilon$ prescription without having to break dual conformal invariance.  

\subsection{The complex integral} \label{toy}
Since the limits of the $v$ integral are independent of $t$, we can perform the $t$ integrals first and then finish with the $v$ integral in \eqref{uvint}.

The complex $(s,t)$ integrals are relatively straightforward  and we first give a toy example, first shown to us by Nima Arkani-Hamed, which we will use later:
\begin{lemma}
\be{}
\int_\Gamma  \frac{\rd s}s\,  \frac{\rd t}t =4\pi i \log \left|\frac ba\right|\quad \mbox{where the contour $\Gamma$ is } \quad s=c\frac{\bar t -a}{\bar t -b}
\label{lem}
\ee
\end{lemma}

This can be seen by expressing the integrand on the contour as the exterior derivative of
\be{}
\log t \,\rd  \log \left( \frac{\bar t -a}{\bar t-b}  \right) =\log t\left( \frac 1{\bar t -a} -\frac 1{\bar t -b}\right )\, \rd \bar t\, \, .
\label{integrand}
\ee
However, this form must have a cut from $t=0$ to $t=\infty$ and furthermore has poles at $\bar t=a$  and $b$.  So in order to use Stokes theorem, we must cut out an $\epsilon$-neighbourhood of the cut and the poles.  We can then use Stokes to reduce the integral to a contour integral around each pole and the cut.  The contour integral around the cut can then be reduced to a line integral along the cut as the contribution from the logarithmic singularities at the end vanishes as $\epsilon\rightarrow 0$, whereas the jump across the cut is $2\pi i$.  The contributions from the poles as $\epsilon\rightarrow 0$ is similarly given as $\mp 2\pi i\log t$ evaluated at the poles (noting the anti-holomorphic dependence on $t$) yielding
\be 
2\pi i\left(\int _0^\infty \rd \log \left( \frac{\bar t -a}{\bar t-b}  \right) - \log {\bar a} +\log {\bar b}\right)= -2\pi i\left(\log \frac{\bar a }{\bar b}+\log \frac {a}{b}\right)  =4\pi i\log \frac{|b|}{|a|} \, . \Box
\ee 

In our context, the complex integral is precisely of this form, but the contour is given by \eqref{eq:real}. With this choice of contour, the lemma for the complex integral gives
\begin{equation}
F(v)=\int_\Gamma  \frac{\rd t}t \, \frac{\rd s}s =4\pi i\ln\left|\frac{\left(a_{i-1j}-v\right)\left(a_{ij-1}-v\right)}{\left(a_{i-1j-1}-v\right)\left(a_{ij}-v\right)}\right|.
\label{Fv}
\end{equation}
As promised, $F(v)\rightarrow 0$ as $v\rightarrow \infty$ and the $v$-integral that we are left with is
\be\label{v-int}
\int_{v_*}^\infty F(v)\rd \ln v=\int_{v_*}^\infty  4\pi i\ln\left|\frac{\left(a_{i-1j}-v\right)\left(a_{ij-1}-v\right)}{\left(a_{i-1j-1}-v\right)\left(a_{ij}-v\right)}\right|\, \rd \ln v.
\ee

\subsection{The answer for generic diagrams}

From the standard definition of the dilog function  we obtain the alternative formula
\be
\Li_2 x=-\int_0^x \log( 1-x')\frac{\rd x'}{x'}=-\int_{1/x}^\infty \log \left(1-\frac 1{x''}\right)\frac{\rd x''}{x''}\, ,
\ee
where we have simply set $x''=1/x'$.  Our integral \eqref{v-int} breaks up into an alternating sum of four such terms 
\be
\int_{v_*}^{\infty}  \log\left(1-\frac {a}{v}\right)  \frac{\rd v}v=-\Li_2 \frac {a}{v_*} 
\, .
\label{li2}
\ee
Hence, the integral in  \eqref{v-int} reduces to 
\begin{equation}
K_{ij}= \Li_{2}\left(\frac{a_{ij}}{v_{*}}\right)+\Li_{2}\left(\frac{a_{i-1j-1}}{v_{*}}\right)-\Li_{2}\left(\frac{a_{i-1j}}{v_{*}}\right)-\Li_{2}\left(\frac{a_{ij-1}}{v_{*}}\right)+c.c.
\label{generic}
\end{equation}

In \cite{Lipstein:2012vs} we verified that \eqref{generic} leads to the correct symbol for the 1-loop MHV amplitude. In particular, we showed that standard expressions for the symbol of the 1-loop MHV amplitude could be expressed as a sum over $i,j$ of
\[
\left(a_{ij-1}a_{i-1j}-a_{ij}a_{i-1j-1}\right)\otimes\frac{a_{ij-1}a_{i-1j}}{a_{ij}a_{i-1j-1}}+\mbox{c.c.}
\]
which isolates those invariants appearing in $K_{ij}$.  This discussion is repeated in appendix \ref{symbol}.  The formula \eqref{generic} above reduces to  this together with the terms 
\begin{equation}\label{boundary}
\Delta_j \left( (a_{ij}-a_{i-1\, j})\otimes \frac{a_{ij}}{a_{i-1\, j}}\right)+\Delta_i\left( (a_{ij}-a_{i\, j-1})\otimes \frac{a_{ij}}{a_{i\, j-1}} \right) +\mbox{c.c.}
\end{equation}
where $\Delta_i f_{ij}=f_{ij}-f_{i-1\, j}$ and $\Delta_j f_{ij}=f_{ij}-f_{ij-1}$ are difference operators, so that the sum of such terms collapses telescopically.

Although the appearance of four dilogs in \eqref{generic} is reminiscent of the four dilogs that were obtained in \cite{Brandhuber:2004yw,Brandhuber:2007yx} for the 1-loop MHV amplitude using the CSW formalism and null polygonal spacetime Wilson loops, respectively, the formulae in these references are quite non-trivially different than ours.  In particular, their formulae are not dual conformal invariant up to the choice of the reference twistor and the reference twistor used in those references is chosen so that its underlying spinor is one of the spinors in a 2-mass-easy box that is related to the Kermit.  As such, in appendix \ref{BST}, where we describe the detailed relations of our formula to those of \cite{Brandhuber:2004yw}, we must work with one of their integral formulae in order to verify our formula against theirs.  

A final remark is that the structure that is at the root of the analysis of \cite{Brandhuber:2004yw}, and the proof in \cite{Bullimore:2010pj} of the equivalence of Kermit integrand to the standard 1-loop MHV integrand, is that Kermit can be expressed as a sum of four one mass boxes (the numerator factor in the original momentum twistor representation of Kermit can be expressed as a sum of 4 terms that each cancels two of the poles leaving boxes \cite{Bullimore:2010pj}). This can be seen directly from the $d\,$log form of the integrand \eqref{dlog} by observing that $s$ and $t$ are natural ratios $s={s_{i-1}}/{s_i}$ and $ t= {t_{j-1}}/{t_j}$,  where 
\be\label{decomp}
 s_{i-1}= \la i-1|x_{0i}|\eta]\, , \quad s_i =\la i|x_{0i}|\eta]
\, ,\quad t_{j-1}=\la j-1|x_{0j}|\eta]\, , \quad t_j=\la j|x_{0j}|\eta]\, ,
\ee
and this gives the decomposition into four terms.  Furthermore, the decomposition into the real part of four dilogs above can be seen to follow from this (although note that as ordinarily expressed, the 1-mass boxes diverge, but the divergent parts cancel between the four terms and is not seen in a generic Kermit).

\section{Divergent Diagrams} \label{divergent}
\begin{figure}[h]
\begin{center}
\includegraphics[scale=0.11]{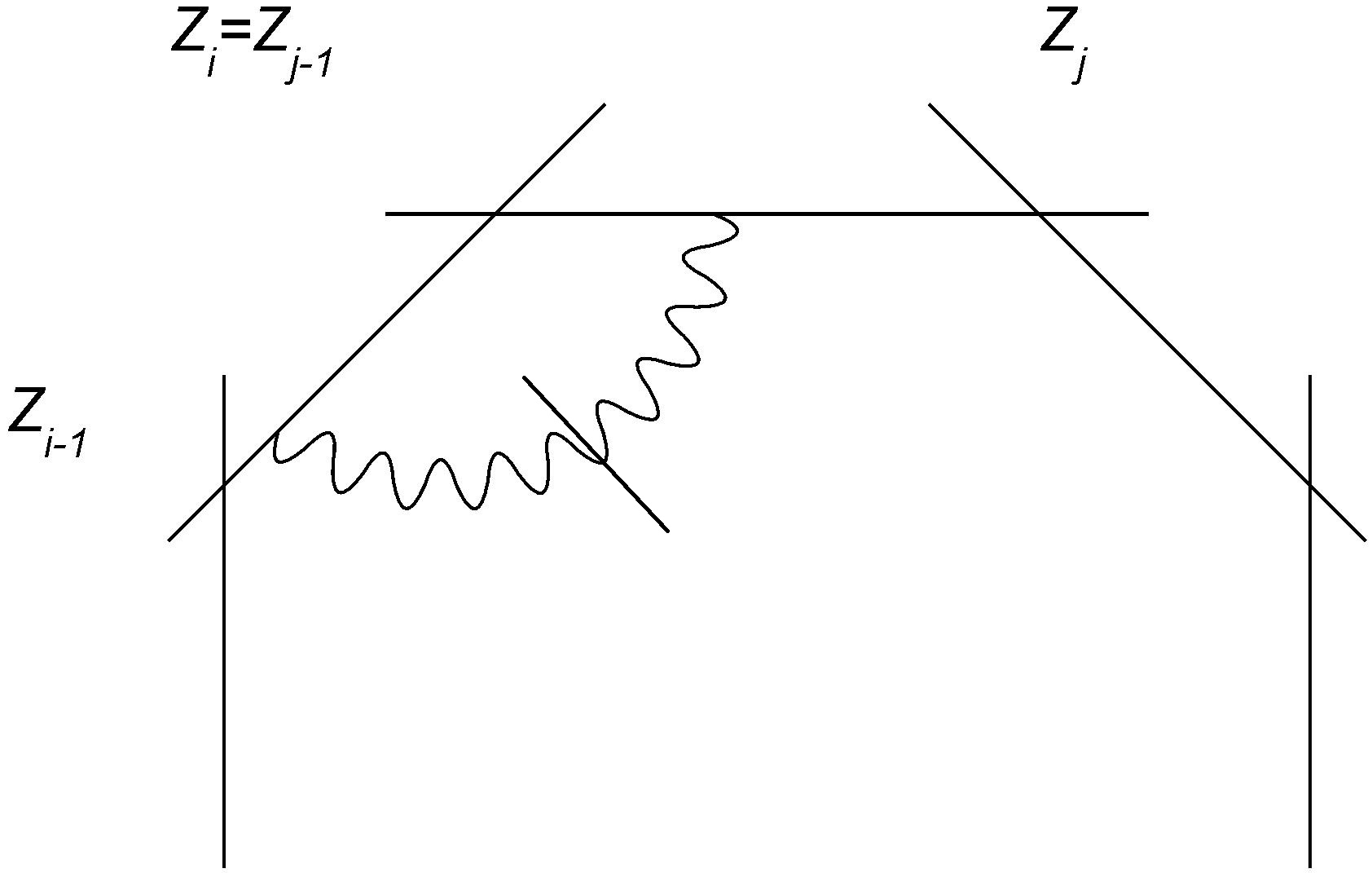}
\caption{A propagator connecting two adjacent edges of the twistor Wilson loop. This diagram diverges.}
\label{div1}
\end{center}
\end{figure}

\begin{figure}[h]
\begin{center}
\includegraphics[scale=0.12]{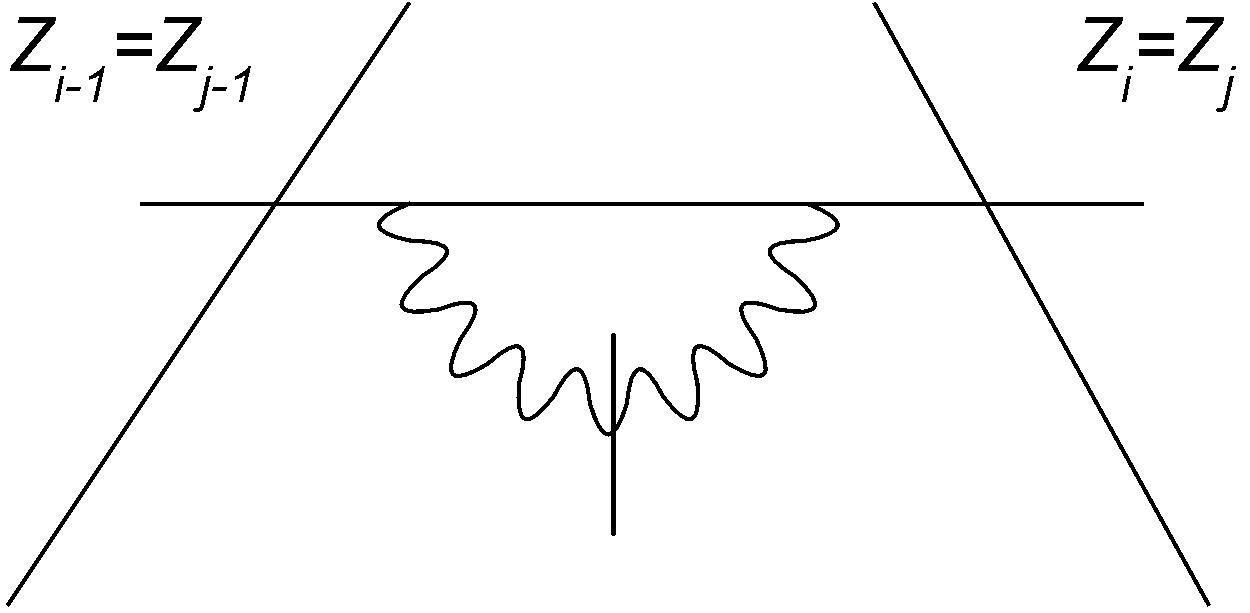}
\caption{A propagator beginning and ending on the same edge of the twistor Wilson loop. This diagram vanishes.}
\label{div2}
\end{center}
\end{figure}

The usual infrared divergence arises from the $v$-integral, which becomes logarithmically divergent when $v_*=0$ and this is then compounded by a divergence in the $t$ integral at $v=0$ to make a $\log^2$ divergence.  If $i=j-2$, then $a_{ij-1}=0$, but $v_*\neq 0$  and the expression for Kermit in \eqref{generic} is still finite. On the other hand, if $i=j-1$, then $a_{ij}=a_{i-1j-1}=a_{ij-1}=0$ so that $v_*=0$ and \eqref{generic} diverges. This corresponds to a Wilson loop diagram in which the propagator connects two adjacent edges, as depicted in Figure \ref{div1}.  If $i=j$, the geometry in our setup of the $d$log form of the integrand breaks down so we must analyze this situation essentially from scratch. This situation corresponds to a Wilson loop diagram in which the propagator begins and ends on the same edge (as depicted in Figure \ref{div2}). Since the dual diagram contains a two-point MHV vertex at one end with no external leg, this diagram must vanish. Hence, the only case that needs to be regulated is $i=j-1$.   
In order to make \eqref{generic} well-defined, we will have to
introduce a regulator. In the following we use mass regulation
\cite{Henn:2011by}.

\subsection{Mass Regularization}
Various simplifications occur when $i=j-1$;  the contour $\Gamma$ simplifies to become
\begin{equation}
s=\frac{\bar{t}\left(a_{i-1i+1}/v-1\right)-1}{\bar{t}+1}.
\label{ijcont} 
\end{equation}
Mass regularization requires that we replace $\left(x_0-x_{i}\right)^{2}\rightarrow\left(x_0-x_{i}\right)^{2}-m^{2}$,
$\left(x_0-x_{j}\right)^{2}\rightarrow\left(x_0-x_{j}\right)^{2}-m^{2}$.
Combining this with the Feynman $i\epsilon$ prescription, we see
that mass regularization is equivalent to taking $i\epsilon\rightarrow i\epsilon-m^{2}$.
  Thus performing this shift in  
\eqref{iepshift} gives
\begin{equation}
\frac1{2\pi i}\int_0^\infty  \rd v \int_\Gamma \frac{\rd t}t \frac{\rd s}s\frac{1}{v+\left(i\epsilon-m^2\right)\left(f_{i}-f_{i+1}\right)}
\label{massdef}
 \end{equation}
where $f_{i},f_{j}$ are as defined above. Here we can no longer simply do the $t$ integral using the lemma above as $f_i$ and $f_j$ depend on $t$.  This is not however too complicated when $i=j-1$, since using $x_{i\, i+1}=\bar \lambda_i\lambda_i$ in equation \ref{fifj}, we have
$$
f_{i}-f_{i+1}=-\frac{1}{2}\frac{\left(v+x_{i\,i+2}^{2}\left|1+t^{-1}\right|^{-2}\right)^{2}}{v \, x_{i\,i+2}^{2}\left|1+t^{-1}\right|^{-2}}.
$$ 

In appendix \ref{massdetails}, we compute the integral in \eqref{massdef} by expanding the integrand in $m$ and neglecting terms of $\mathcal{O}(m)$. In the end, we obtain  
\begin{equation}
-\frac{1}{4}\left(\ln^{2}\left(\frac{m^{2}}{x_{ii+2}^{2}}\right)+\ln\left(\frac{x_{i-1i+1}^{2}}{x_{ii+2}^{2}}\right)\ln\left(x_{ii+2}^{2}\right)\right)-\frac{2\pi^2}{3}+\mathcal{O}(m).
\label{massresult1}
\end{equation}
In appendix \ref{massdetails}, we also show that this expression reproduces the form of the 4-point 1-loop MHV amplitude obtained in \cite{Alday:2009zm}.

\section{Higher Loops} \label{higherloops}

Our calculation of the 1-loop MHV amplitude from Kermit is in fact already a significant first step in the computation of  higher loop amplitudes. As a start, at two loops, every diagram has at least one Kermit as a subdiagram and one can perform at least half the integrals directly and this will be true of the majority of diagrams at higher loop order.   In general, it follows from the construction of the $d\,$log form given in \cite{Lipstein:2012vs} that it will always be possible to associate two propagators to each vertex and obtain two real and two complex $d\, $log  integrals as for Kermit.  The two real integrals will then need to be regulated via an $i\epsilon$ prescription that will follow analogously to our arguments given here.

\begin{figure}[h]
\begin{center}
\includegraphics[scale=0.15]{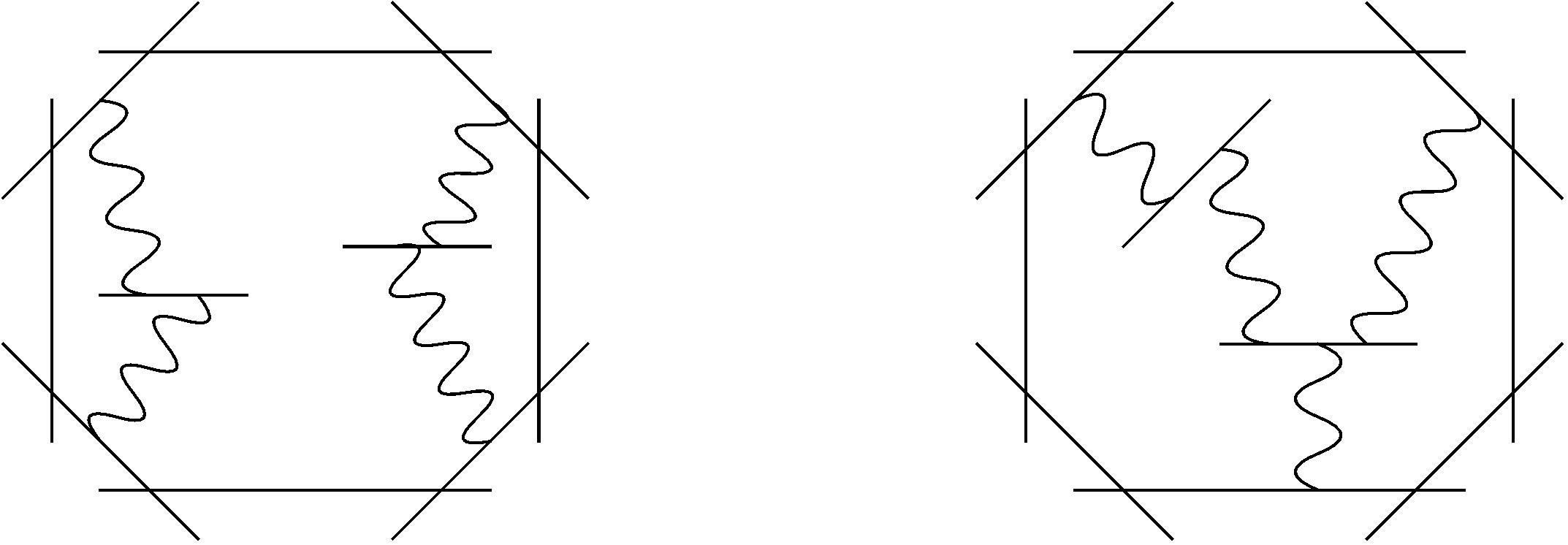}
\caption{The two topologies of nonzero Wilson loop diagrams contributing to the 2-loop MHV amplitude.}
\label{fig2loop}
\end{center}
\end{figure}

The general story at two-loops is still relatively simple and we sketch the calculation of a non-trivial two-loop MHV-diagram using the techniques described in this paper. As described in \cite{Lipstein:2012vs}, there are only two topologies of nonzero twistor Wilson loop diagrams corresponding to 2-loop MHV amplitudes. These are illustrated in Figure \ref{fig2loop}. There are also boundary versions of these diagrams, where one or two pairs of propagators end on the same edge of the Wilson loop. From these, we can see that Kermit is in some sense a basic building block for higher-loop loop amplitudes. For example, the diagram on the left in Figure \ref{fig2loop} is simply given by the product of two Kermits and therfore has the form $(\Li_2)^2$.    

\begin{figure}[h]
\begin{center}
\includegraphics[scale=0.2]{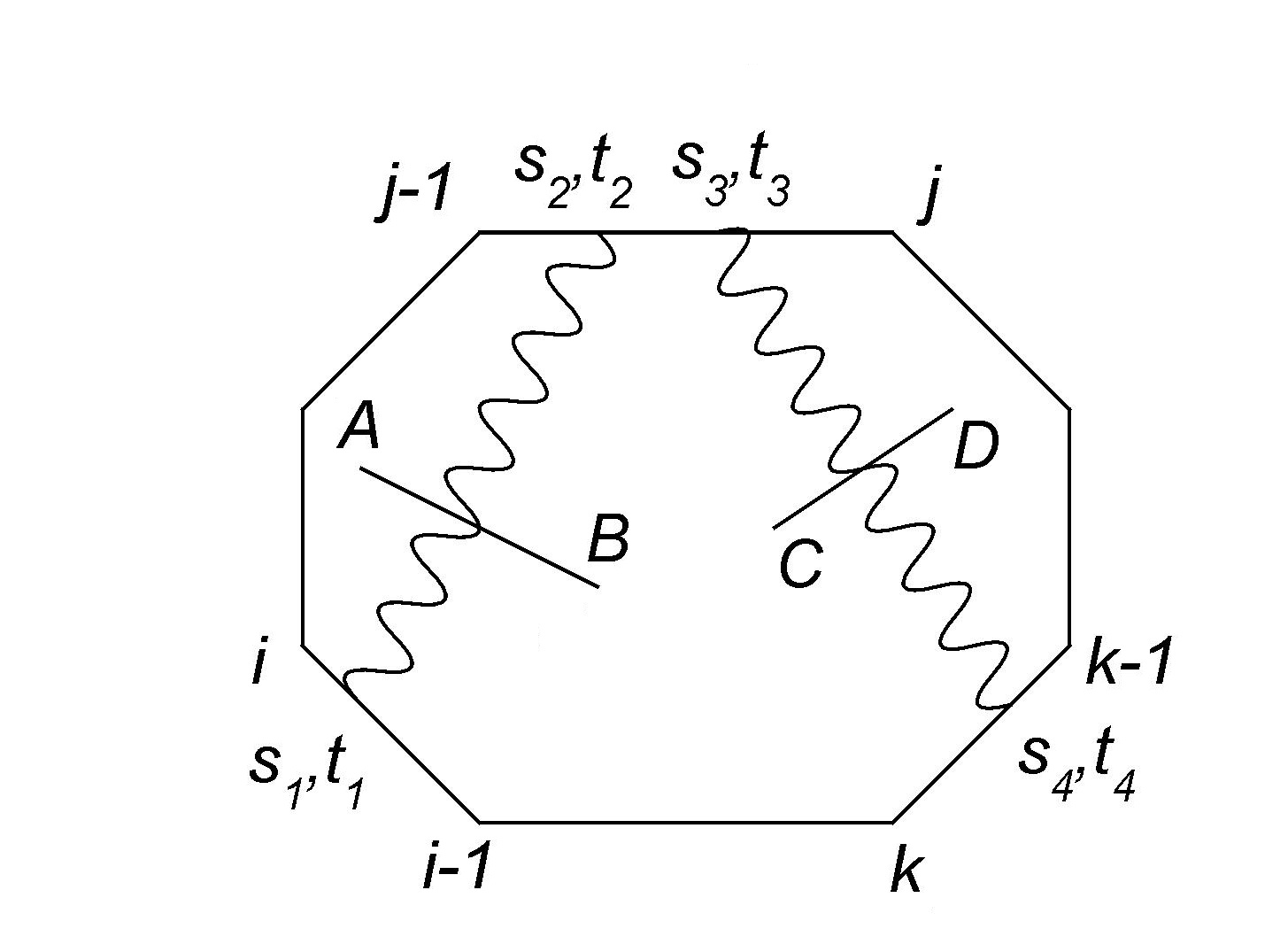}
\caption{Boundary version of a Wilson loop diagram contributing to the 2-loop MHV amplitude.}
\label{fig1}
\end{center}
\end{figure}

A less trivial case is illustrated in Figure \ref{fig1}. In this case, the amplitude does not factorize into a product of two Kermits, and we will see that one obtains also polylogs of higher degree such as $\Li_4$.  
The integral for the diagram in Figure \ref{fig1} is
\[
\int\frac{ds_{1}}{s_{1}}\frac{dt_{1}}{t_{1}}\frac{ds_{2}}{s_{2}}\frac{dt_{2}}{t_{2}}\frac{ds_{3}}{s_{3}-s_{2}}\frac{dt_{3}}{t_{3}}\frac{ds_{4}}{s_{4}}\frac{dt_{4}}{t_{4}}\]
where the momentum twistors are parametrized as follows:
\[
Z_{A}=Z_{*}+t_{1}\left(Z_{i-1}+s_{1}Z_{i}\right),\,\,\, Z_{B}=Z_{*}+t_{2}\left(Z_{j-1}+s_{2}Z_{j}\right)\]
\[
Z_{C}=Z_{*}+t_{3}\left(Z_{j-1}+s_{3}Z_{j}\right),\,\,\, Z_{D}=Z_{*}+t_{4}\left(Z_{k-1}+s_{4}Z_{k}\right).\]
In this case, we can treat the right-hand Kermit in Figure \ref{fig1} as a standard Kermit, but one in which the fixed twistors are $(Z_{j-1}+s_2Z_j,Z_j)$, and $(Z_{k-1},Z_k)$.  This factor can be done first leading to dilogs depending on $s_2$.  To implement this, we perform the following change of variables (which incorporates those we used before to make the real poles apparent):
\[
\left(s_{1},t_{1}\right)=\left(-s,-\frac{i}{s_{0}\left(1+s\right)}\right),\,\,\,\left(s_{2},t_{2}\right)=\left(t,-\frac{i}{t_{0}\left(1+t\right)}\right)\]
\[
\left(s_{3},t_{3}\right)=\left(u+t,-\frac{i}{u_{0}\left(1+u+t\right)}\right),\,\,\,\left(s_{4},t_{4}\right)=\left(-w,-\frac{i}{w_{0}\left(1+w\right)}\right).\]
In terms of the new variables, the integral becomes
\begin{equation}
\int 
 \rd \ln s_{0} \rd \ln t_{0}  \rd \ln t \rd \ln s \int \rd \ln u_{0} \rd \ln w_{0}  \rd \ln u \rd \ln w 
\label{eq:2loop}\end{equation}
where we have split the 2-loop integral into two factors, each of which is now very similar to Kermit. In particular, the contour is determined by:
\[
s_{0}=\bar{s}_{0},\,\,\, t_{0}=\bar{t}_{0},\,\,\, w_{0}=\bar{w}_{0},\,\,\, u_{0}=\bar{u}_{0}\]
\begin{equation}
s=\frac{\bar{t}\left(a_{i-1j}+v_{-}\right)+a_{i-1j-1}+v_{-}}{\bar{t}\left(a_{ij}+v_{-}\right)+a_{ij-1}+v_{-}}
\label{scons}
\end{equation}
\begin{equation}
w=\frac{\left(\bar{u}+\bar{t}\right)\left(a_{k-1j}+y_{-}\right)+a_{k-1j-1}+y_{-}}{\left(\bar{u}+\bar{t}\right)\left(a_{kj}+y_{-}\right)+a_{kj-1}+y_{-}}
\label{wcons}
\end{equation}
where \begin{equation}
v_{\pm}=s_{0}\pm t_{0},\,\,\, y_{\pm}=w_{0}\pm u_{0}.\label{eq:vpmypm}\end{equation}

\begin{figure}[h]
\begin{center}
\includegraphics[scale=0.2]{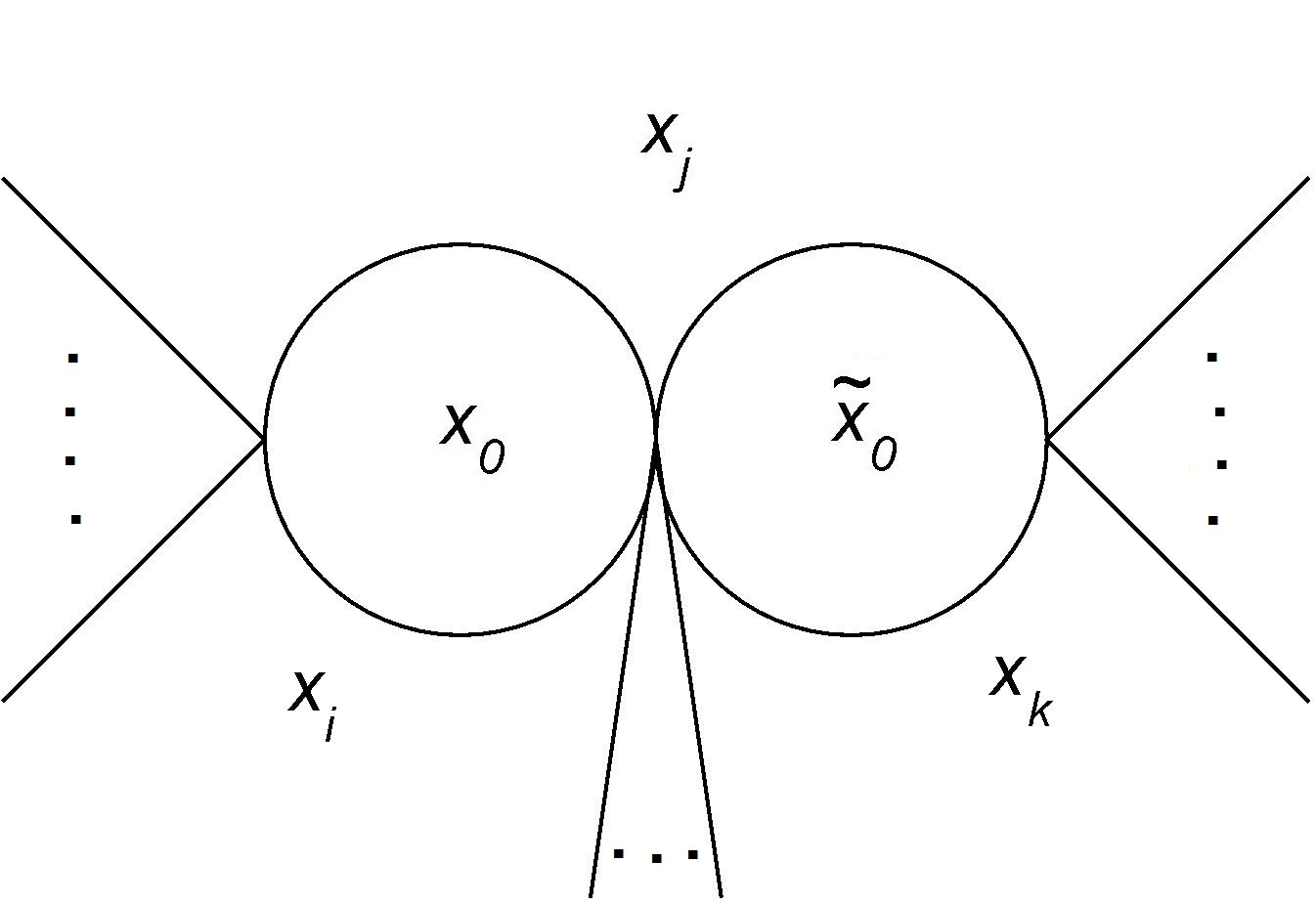}
\caption{The CSW diagram corresponding to the Wilson loop diagram in Figure \ref{fig1}.}
\label{fig2}
\end{center}
\end{figure}

The $u_0,w_0,u,w$ integrals in \eqref{eq:2loop} can now be evaluated exactly as in the evaluation of Kermit. This will yield a sum of dilogs which now depends on the variables $(t,\bar{t})$, since $\bar{t}$ appears on the definition of the contour of the the $u,w$ integral in \eqref{wcons}. Hence, the $s_0,t_0,s,t$ integrals in \eqref{eq:2loop} will be like those in Kermit, but with dilogs in the integrand. In more detail, the integral in \eqref{eq:2loop} has real poles in the variables $s_0,t_0,w_{0},u_{0}$ and we must make these well-defined using the Feynman $i\epsilon$ prescription.  In terms of the the $v_{\pm},y_{\pm}$ variables defined in \eqref{eq:vpmypm} we have: 
\begin{equation}
\int\frac{2\rd v_{+}\rd v_{-}}{\left(v_{+}+v_{-}\right)\left(v_{+}-v_{-}\right)}\int_{\Gamma_s} \rd \ln t \rd \ln s
\int\frac{2\rd y_{+}\rd y_{-}}{\left(y_{+}+y_{-}\right)\left(y_{+}-y_{-}\right)}\int_{\Gamma_w} \rd \ln u \rd \ln w\label{eq:2loop2}\end{equation}
where $\Gamma_s$ refers to the contour defined by \eqref{scons} and $\Gamma_w$ by \eqref{wcons}.
The real variables $\left(s_{0},t_{0},w_{0},u_{0}\right)$ are related to the region momenta by
\[
s_{0}=\frac{x_{0i}^{2}}{2\left[\eta\right|x_{0i}\left|\bar{\eta}\right\rangle },\,\,\, t_{0}=\frac{x_{0j}^{2}}{2\left[\eta\right|x_{0j}\left|\bar{\eta}\right\rangle },\,\,\, w_{0}=\frac{x_{\tilde{0}k}^{2}}{2\left[\eta\right|x_{\tilde{0}k}\left|\bar{\eta}\right\rangle },\,\,\, u_{0}=\frac{x_{\tilde{0}j}^{2}}{2\left[\eta\right|x_{\tilde{0}j}\left|\bar{\eta}\right\rangle }\]
where the region momenta are depicted in Figure \ref{fig2}. 

The $i\epsilon$
prescription then amounts to
\[
v_{\pm}\rightarrow v_{\pm}+i\epsilon\left(f_{i}\pm f_{j}\right),\,\,\, y_{\pm}\rightarrow y_{\pm}+i\epsilon\left(\tilde{f}_{k}\pm\tilde{f}_{j}\right)\]
where
\[
f_{i}=\frac{1}{2x_{0i}\cdot\eta},\,\,\, f_{j}=\frac{1}{2x_{0j}\cdot\eta},\,\,\,\tilde{f}_{k}=\frac{1}{2x_{\tilde{0}k}\cdot\eta},\,\,\,\tilde{f}_{j}=\frac{1}{2x_{\tilde{0}j}\cdot\eta}.\]
We can now  evaluate the $\left(v_{+},y_{+}\right)$ integrals in \eqref{eq:2loop2} as before to obtain
\begin{equation}
-4\pi^{2}\int_{v_{*}}^{\infty} \rd \ln v_{-}\int_{\Gamma_s} \rd \ln t\rd \ln s\int_{y_{*}}^{\infty}\rd \ln y_{-}\int_{\Gamma_w} \rd \ln u\rd\ln w
\label{2loop3}
\end{equation}
where 
\[
v_{*}=-\frac{x_{ij}^{2}}{2x_{ij}\cdot\eta},\,\,\, y_{*}=-\frac{x_{jk}^{2}}{2x_{kj}\cdot\eta}.
\] 
We now evaluate the $(u,w)$ integral using \eqref{lem} and the $y_{-}$ integral using \eqref{li2} to obtain
\begin{multline}
\int_{y_{*}}^{\infty}\rd\ln y_{-}\int_{\Gamma_w} \rd\ln u\rd\ln w=
-2\pi i\left[\Li_{2}\left(\frac{a_{k-1j}}{y_{*}}\right)-\Li_{2}\left(\frac{a_{kj}}{y_{*}}\right)\right. \\ \left. +\Li_{2}\left(\frac{\left(\bar{t}a_{kj}+a_{kj-1}\right)}{\left(\bar{t}+1\right)y_{*}}\right)-\Li_{2}\left(\frac{\left(\bar{t}a_{k-1j}+a_{k-1j-1}\right)}{\left(\bar{t}+1\right)y_{*}}\right)+c.c.\right].
\label{uwy}
\end{multline}
Since the first two dilogs do not depend on $t$ or $\bar{t}$, when we perform the $(v_{-},t,s)$ integrals in \eqref{2loop3} against these terms, the computation will be identical to that of Kermit, and we will be left with terms of form $\left(\Li_{2}\right)^{2}$. On the other hand, integrating $(v_{-},t,s)$ against the second two terms in \eqref{uwy} and their complex conjugates is nontrivial and will yield $\Li_4$'s. For example, we have integrals of the form
\begin{equation}
\int_{v_{*}}^{\infty}\rd\ln v_{-}\int_{\Gamma} \rd \ln s\rd \ln t\, \Li_{2}\left(\frac{at+b}{t+1}\right)
\label{li4}
\end{equation}
where $(a,b)$ are constants and the contour $\Gamma$ is defined by the reality constraint
\[
 s=\frac{\bar{t}-c(v_{-})}{\bar{t}-d(v_{-})},
\]
where $(c(v_{-}),d(v_{-}))$  are Mobius transformations of $v_-$ following from \eqref{scons}. Writing
\[
\int_{\Gamma} \rd \ln s\rd\ln t \,\Li_{2}\left(\frac{at+b}{t+1}\right)=\int_{\Gamma}\rd\left( \ln s  \,\Li_{2}\left(\frac{at+b}{t+1}\right) \rd \ln t\right),
\]
we see that the integral can be evaluated using Stokes theorem taking into account the singularities. We find
\begin{multline}
\int_{\Gamma} \rd \ln s \rd \ln t \,\Li_{2}\left(\frac{at+b}{t+1}\right)=2\pi i\left[\ln\left(\frac{c(v_{-})}{d(v_{-})}\right)\Li_{2}(b)+\int_{\bar{c}(v_{-})}^{\bar{d}(v_{-})}\rd\ln t \;\Li_{2}\left(\frac{at+b}{t+1}\right)\right.
\\
\left.+(b-a)\int_{1}^{\infty}\frac{\rd z}{\left(a-z\right)\left(b-z\right)}\ln\left(\frac{\left(c(v_{-})+1\right)z-c(v_{-})a-b}{\left(d(v_{-})+1\right)z-d(v_{-})a-b}\right)\ln z\right]\, .
\label{li42}
\end{multline}
The first term on the right comes from the pole at $t=0$ (the residue of the pole at $t=\infty$ vanishes), the second  from the branch cut of $\ln s$ (which runs from $s=0$ to $s=\infty$), and the third from the branch cut of the dilog in \eqref{li4} ($\Li_2(z)$ has a branch cut along the positive $z$ axis running from $z=1$ to $z=\infty$ with discontinuity $2\pi i \ln|z|$).
    
When the integral over $v_-$ in \eqref{li4} is performed against the first term on the right hand side of \eqref{li42}, one obtains functions of the form $(\ln)^2 \times Li_{2}$ and $\left(Li_{2}\right)^2$. Furthermore, performing the $t$ and $z$ integrals in the second and third terms on the right hand side of \eqref{li42} will yield terms of transcendentality 3, notably terms of the form $(\ln )^3$, $\ln \times  \Li_2 $, and $\Li_3$. Peforming the $v_{-}$ integral against these terms then yields functions of transcendentality 4 including $Li_4$.   

\section{Conclusion} \label{conclusion}

We have seen how to compute certain scattering amplitudes and
correlation functions in $\mathcal{N}=4$ sYM without Feynman
parameters. The starting point is to express the loop integrands in
$d\, \log$  form. In our approach this is automatic for the MHV part
of the holomorphic Wilson loop in momentum twistor space to all loop orders, whose Feynman rules
can be written $d\, \log$  form \cite{Lipstein:2012vs}. An identical
approach will work for the twistor space formulation of amplitudes as
given in \cite{Adamo:2011cb} which gives formulae for amplitudes that
are remarkably similar to the formulae for the Wilson loop given
here. The integration variables in the $d\, \log$  form have a very
simple geometric interpretation: they correspond to insertion points
of propagators onto MHV vertices or edges of the Wilson loop (and for
the amplitude in ordinary twistor space, they will be insertion points
on just the MHV vertices).   An alternative, and generically
inequivalent  $d\, \log$  form for the planar amplitudes of
$\mathcal{N}=4$ sYM was given using on-shell diagrams and the
Grassmannian integral formula  in \cite{ArkaniHamed:2012nw} (although
this is essentially equivalent at 1-loop MHV).  Our methods should
apply there also if one can identify the real poles corresponding to
physical propagators from amongst the unphysical complex poles.

The external data on which the answers depend are encoded in the
integration contours that arise from reality constraints in momentum
twistor space. The main subtlety that arises is the appropriate
regularisation for the real poles that arise. Since these are
associated with physical propagators,  the standard Feynman
$i\epsilon$ prescription must be used. This can be done explicitly and
then the integrals become well-defined, and can be evaluated using
Stokes theorem. 
In the non-generic divergent case, mass regularization can be
implemented by including a mass together with the Feynman $i
\epsilon$. 

This procedure leads to a remarkably simple expression for the generic
`Kermit' diagram which remarkably turns out to be dual conformal invariant. Given that the Feynman $i\epsilon$ prescription breaks dual conformal invariance, this did not need to be the case.  This form of the 1-loop MHV amplitude is nontrivially related to standard ones in the literature, as we demonstrate in Appendix \ref{BSTa}.  Our method for computing the planar 1-loop MHV amplitude appears to be both direct and simpler than previous methods.  Furthermore, we provide evidence in section \ref{higherloops} that this method will scale  to higher loop amplitudes because Kermit is a  basic building block for many higher loop diagrams for the twistor Wilson loop. It seems potentially tractable  to use the techniques described here to carry out a complete evaluation of the 2-loop MHV amplitudes and compare this with the results with \cite{CaronHuot:2011ky,Golden:2013lha}, which obtained the differential of all planar 2-loop MHV amplitudes.  The Regge limit of higher loop MHV amplitudes might also provide a useful testing ground that would allow easier evaluation to compare with the results of \cite{Lipatov:2010qg,Bartels:2011ge,Dixon:2012yy}.

We believe that these methods can be used to compute loop amplitudes or correlation functions more generally.  In order to obtain the $d\,$log form we did not need to assume that the diagrams were planar when computing the holomorphic Wilson loop.  For the twistor space computation of amplitudes in \cite{Adamo:2011cb}, planarity was again not assumed, and indeed the basic framework extends to different amounts of supersymmetry.  The straightforward  $d\,$log form did seem however to require the assumption that the amplitude be MHV, and so extending the ideas beyond MHV remain a challenge. A different extension given in \cite{ArkaniHamed:2012nw} proposed an on-shell diagram formalism for the ABJM theory (a $\mathcal{N}=6$ superconformal Chern-Simons theory in three dimensions \cite{Aharony:2008ug}), and suggested that the loop integrands of the amplitudes of this theory should also have a $d\, \log$  form. In \cite{ArkaniHamed:2012nw}, they also suggest that the on-shell diagram formalism can be extended to 4d super Yang-Mills theories with $1\leq \mathcal{N}\leq4$, so it is conceivable that the integrands of loop amplitudes of these theories can also be expressed in $d\, \log$  form and so these are directions for further investigation.

\section*{Acknowledgments}
We would like to thank Nima Arkani-Hamed, Zvi Bern, Simon Caron-Huot, Lance Dixon, Gregory Korchemsky, and David Skinner for helpful discussions.  AL is supported by a Simons Postdoctoral Fellowship; LM is supported by a Leverhulme Fellowship and EPSRC grant number EP/J019518/1. 



\appendix

\section{Comparison to Known Results} \label{BSTa}

In this appendix, we will relate our result for the generic Kermit diagram in \eqref{generic} to the results of \cite{Brandhuber:2004yw}. The planar 1-loop MHV amplitude for $\mathcal{N}=4$ sYM was computed in \cite{Brandhuber:2004yw} from MHV diagrams using the unitarity cut method. They found that the MHV diagram in Figure \ref{1lmhv} can be separated into four pieces of the form:
\begin{equation}
\frac{1}{\epsilon}\int\frac{dz}{z}\left|\frac{P_{L;z}^{2}}{4}\right|^{-\epsilon}\left(1-a_{z}P_{L;z}^{2}\right)^{\epsilon}+\mathcal{O}\left(\epsilon\right)\label{eq:disc}\end{equation}
where $\epsilon$ is the dimensional regularization parameter and
\begin{equation}
P_{L;z}=x_{ij}-z\eta,\,\,\, a_{z}=\frac{p_{m_{1}}\cdot p_{m_{2}}}{N\left(P_{L;z}\right)},\,\,\, N\left(P\right)=-\frac{1}{2}\left|\left\langle m_{1}\right|P\left|m_{2}\right]\right|^{2},\label{eq:params}\end{equation}
where $\eta$ is a reference momentum and $\left(m_{1},m_{2}\right)$ label two external legs. Kermit is obtained by summing over $\left(m_{1},m_{2}\right)=(i,j),(i-1,j-1)$ and subtracting $\left(m_{1},m_{2}\right)=(i-1,j),(i,j-1)$. We have
\[
P_{L;z}^{2}=2 n\cdot x_{ij}\left(v_{*}-z\right)\]
where $n= \bar{\eta} \eta $ and $v_{*}$ corresponds to
\[
v_{*}=\frac{x_{ij}^{2}}{2n\cdot x_{ij}}
.\] 
Since the integral over $z$
in \eqref{eq:disc} is such that $P_{L;z}^{2}\geq0$, this implies
that the limits of the integral are $v_{*}\leq z\leq\infty$ (assuming $2\eta\cdot x_{ij}<0$). Expanding
$\left(1-a_{z}P_{L;z}^{2}\right)^{\epsilon}$ in the integrand of
\eqref{eq:disc} to $\mathcal{O}(\epsilon)$ then gives \begin{equation}
\int_{v_{*}}^{\infty}\frac{dz}{z}\left(\frac{1}{\epsilon}\left|\frac{P_{L;z}^2}{4}\right|^{-\epsilon}+\log\left(1-a_{z}P_{L;z}^{2}\right)\right)+\mathcal{O}\left(\epsilon\right).\label{eq:disc2}\end{equation}

Recall that we defined the inner product between two external external twistors to be $a_{m_{1}m_{2}}=iZ_{m_{1}}\cdot\bar{Z}_{m_{2}}$. In terms of spinors, this is given by 
\begin{equation}
a_{m_{1}m_{2}}=\frac{\left\langle m_{1}\right|x_{ij}\left|m_{2}\right ] }{\left\langle m_{1} \bar{\eta}\right\rangle \left[\eta m_{2}\right]}.
\label{amn}
\end{equation}
Plugging eqs \ref{eq:disc} and \ref{amn} into \ref{eq:disc2} we obtain 
\[
\int_{v_{*}}^{\infty}\frac{dz}{z}\ln\left(1+\frac{p_{m_{1}}\cdot p_{m_{2}}\left(x_{ij}^{2}-2zx_{ij}\cdot n\right)}{2p_{m_{1}}\cdot n p_{m_{2}}\cdot n\left|a_{m_{1}m_{2}}+iz\right|^{2}}\right)\]
where we have discarded the first (divergent) term in the integrand of \ref{eq:disc2},
since it cancels out among the four contributions to Kermit. Writing
the above equation as
\begin{equation}
\int_{v_{*}}^{\infty}\frac{dz}{z}\left[-\log\left|a_{m_{1}m_{2}}-z\right|^{2}+\log\left(\left|a_{m_{1}m_{2}}-z\right|^{2}+4p_{m_{1}}\cdot p_{m_{2}} n\cdot x_{ij}\left(v_{*}-z\right)\right)\right]
\label{BST}
\end{equation}
we see that the first term has the same form the integral we obtained in \eqref{v-int} and gives our result for Kermit after summing $\left(m_{1},m_{2}\right)=(i,j),(i-1,j-1)$ and subtracting $\left(m_{1},m_{2}\right)=(i-1,j),(i,j-1)$. Let's focus on the second term.

The integral of the second term in \eqref{BST} is given by: 
\begin{equation}
\int_{v_{*}}^{\infty}\frac{dz}{z}\ln\left[ z^{2}+\left(i\left(a_{m_{2}m_{1}}-a_{m_{1}m_{2}}\right)-4p_{m_{1}}\cdot p_{m_{2}} n \cdot x_{ij}\right)z+\left|a_{m_{1}m_{2}}\right|^{2}+4v_{*}p_{m_{1}}\cdot p_{m_{2}} n \cdot x_{ij}\right].
\label{disc}
\end{equation}
Note that the polynomial can be expressed in factorized form as follows:
\[
\left(z-\frac{p_{m_{1}}\cdot x_{ij}}{p_{m_{1}}\cdot n}\right)\left(z-\frac{p_{m_{2}}\cdot x_{ij}}{p_{m_{2}}\cdot n}\right).
\]
Hence, \eqref{disc} reduces to
\[
\int_{v_{*}}^{\infty}\frac{dz}{z}\left[\ln\left(z-\frac{p_{m_{1}}\cdot x_{ij}}{p_{m_{1}}\cdot n}\right)+\ln\left(z-\frac{p_{m_{2}}\cdot x_{ij}}{p_{m_{2}}\cdot n}\right)\right]
\]
which clearly vanishes after summing over $\left(m_{1},m_{2}\right)=(i,j),(i-1,j-1)$
and subtracting $\left(m_{1},m_{2}\right)=(i-1,j),(i,j-1)$. 
Thus, the integral we obtained in \eqref{v-int} is equivalent (although nontrivially so) to that obtained in \cite{Brandhuber:2004yw}.

\section{Details of Mass Regularization} \label{massdetails}

We rescale 
\begin{equation} 
\tilde{m}^2=\frac{m^2}{x_{i\, i+2}^2}\, , \quad v\rightarrow x_{i\, i+2}^{2}v, \quad  \mbox{ and } \quad \tilde{a}=\frac{a_{i-1 i+1}}{x_{i\, i+2}^{2}}=\frac{\la i-1\, i\ra}{\la i\, i+1\ra}.  
\label{masstilde}
\end{equation}
Setting $\epsilon=0$ in \eqref{massdef} gives
\begin{equation}
\frac1{2\pi i}\int_{0}^{\infty}dv\int_{\Gamma}d\ln td\ln s\left(v+\tilde{m}^{2}\frac{\left(v+\left|1+t^{-1}\right|^{-2}\right)^{2}}{v\left|1+t^{-1}\right|^{-2}}\right)^{-1}
\label{massdef2}
\end{equation}
The contour is defined by
\begin{equation}
s=\frac{\bar{t}\left(\tilde{a}/v-1\right)-1}{\bar{t}+1}\, .
\label{ijcont2} 
\end{equation} 

If we set $\alpha(t)=|1+t^{-1}|^{-1}$,  then \eqref{massdef2} becomes
\begin{equation}\label{massdef1}
\frac1{2\pi i}\int_0^\infty \int_\Gamma \rd v \frac{\rd t}t \frac{\rd s}s\frac{v\alpha^2}{v^2\alpha^2+\tilde{m}^2 (v+\alpha^2)^2}.
\end{equation} 
Note that
\begin{equation}
\frac{v\alpha^{2}}{v^{2}\alpha^{2}+\tilde{m}^{2}\left(v+\alpha^{2}\right)^{2}}=\frac{\alpha}{2}\left(\frac{1}{v\alpha+i\tilde{m}\left(v+\alpha^{2}\right)}+c.c.\right). \label{eq:partial}\end{equation}
Plugging eqs \ref{ijcont2} and \ref{eq:partial} into \eqref{massdef1} gives 
\begin{equation}
\frac{\tilde{a}}{4\pi i}\int\frac{d\bar{t}}{\bar{t}+1}\frac{dt}{t}\alpha\int_{0}^{\infty}\frac{dv}{\bar{t}\tilde{a}-\left(\bar{t}+1\right)v}\left(\frac{1}{\left(\alpha+i\tilde{m}\right)v+i\tilde{m}\alpha^{2}}+\tilde{m}\rightarrow-\tilde{m}\right).\label{eq:mi2}\end{equation}
The $v$ integral in \eqref{eq:mi2} can be evaluated using the formula
\[
\int_{0}^{\infty}\frac{dv}{\left(av+b\right)\left(cv+d\right)}=\frac{1}{ad-bc}\ln\left(\frac{ad}{bc}\right).\]
Performing the $v$ integral in \eqref{eq:mi2} gives
\begin{equation}
-\frac{1}{4\pi i}\int\frac{d\bar{t}}{\bar{t}+1}\frac{dt}{t}\alpha\left[\frac{1}{\frac{i\tilde{m}}{\tilde{a}}\alpha^{2}\left(\bar{t}+1\right)+\left(\alpha+i\tilde{m}\right)\bar{t}}\ln\left(-\frac{i\tilde{m}\alpha^{2}\left(\bar{t}+1\right)}{\left(\alpha+i\tilde{m}\right)\tilde{a}\bar{t}}\right)+\tilde{m}\rightarrow-\tilde{m}\right].
\label{massreg10}
\end{equation} 
If we now perform the following change of variables
\[
q=\left(\bar{t}^{-1}+1\right)/\tilde{a}
\]
the integral in \eqref{massreg10} becomes
\[
-\frac{1}{4\pi i}\int\frac{dqd\bar{q}}{q\left(\bar{q}-1/\bar{\tilde{a}}\right)}\left[\left(1+\frac{i\tilde{m}|q|\left(|\tilde{a}|^{2}+1/\bar{q}\right)}{|\tilde{a}|}\right)^{-1}\ln\left(-\frac{i\tilde{m}|q|}{|\tilde{a}|\bar{q}\left(1+i\tilde{m}|\tilde{a}q|\right)}\right)+\tilde{m}\rightarrow-\tilde{m}\right].
\]
Expanding the denominator of the loagarithm in $\tilde{m}$ and setting $q=r e^{i \theta}$ leads to further simplifications:
\[
\frac{1}{4\pi}\int\frac{drd\theta}{r-e^{i\theta}/\bar{\tilde{a}}}\left[\left(1+\frac{i\tilde{m}}{|\tilde{a}|}\left(|\tilde{a}|^{2}r+e^{i\theta}\right)\right)^{-1}\ln\left(-\frac{i\tilde{m}}{|\tilde{a}|}e^{i\theta}\right)+\tilde{m}\rightarrow-\tilde{m}\right]+\mathcal{O}\left(\tilde{m}\right).  
\]
Performing the integral over polar coordinates then gives 
\begin{equation}
-\frac{1}{4}\ln\tilde{m}^{2}+\frac{1}{4}\ln^{2}|\tilde{a}|^{2}-\frac{1}{2}\ln\bar{\tilde{a}}\left(\ln|\tilde{a}|^{2}+\ln x_{ii+2}^{2}\right)-\frac{2\pi^{2}}{3}+\mathcal{O}\left(\tilde{m}\right).
\label{divker1}
\end{equation}
In obtaining this result, we noted that $\tilde{a}=\frac{\left\langle i-1i\right\rangle }{\left\langle ii+1\right\rangle }$, so terms which are linear in $\ln{\tilde{a}}$ or $\ln\bar{\tilde{a}}$ cancel out telescopically when we sum over all pairs of adjacent edges of the Wilson loop. Furthermore, this sum will also give a contribution corresponding to the complex conjugate of \eqref{divker1}, so we can replace $\ln\bar{\tilde{a}}\rightarrow\frac{1}{2}\ln|\tilde{a}|^{2}$ in that equation. We are then left with
\begin{equation}
-\frac{1}{4}\left(\ln^{2}\left(\frac{m^{2}}{x_{ii+2}^{2}}\right)+\ln\left(\frac{x_{i-1i+1}^{2}}{x_{ii+2}^{2}}\right)\ln\left(x_{ii+2}^{2}\right)\right)-\frac{2\pi^2}{3}+\mathcal{O}(m),
\label{divker2}
\end{equation} 
where we plugged in the definitions of $\tilde{m}$ and $\tilde{a}$.

For the case of four external legs,  when \eqref{divker2} is summed over all pairs of adjacent edges we obtain
\[
-\ln^{2}\frac{m^{2}}{s}-\ln\frac{m^{2}}{t}+\ln^{2}\frac{s}{t}-\frac{16\pi^{2}}{3}=-2\ln\frac{m^{2}}{s}\ln\frac{m^{2}}{t}-\frac{16\pi^{2}}{3}
\]  
where $s$ and $t$ are Mandelstam variables. This agrees with the result for the 4-point 1-loop MHV amplitude obtained in \cite{Alday:2009zm} up to the constant term.

\section{The Symbol of the 1-loop MHV Amplitude} \label{symbol}
According to \cite{CaronHuot:2011ky}, the symbol of the remainder term in the  MHV amplitude is 
$$
d R_n= \sum_{ij} \log u_{i\, j-1\, j\, i-1} \rd \log (i-1\, i \, i+1\,j) \qquad u_{i\, j-1\, j\, i-1}= \frac{x_{i\,j-1}^2 x^2_{j\,i-1}}{x_{ij}^2x^2_{i-1\, j-1}}\,
$$
Kermit can only know about the lines $X_i$ and $X_j$ but up to normalisation $(i-1\, i\, i+1 j)=a_{ji}$.  So decomposing and resumming, we can decompose into the  parts of the symbol  Kermit must provide as follows
\be\label{MHV-symb}
d R_n= \sum_{ij} \log (i-1\, i\, j-1\, j) \rd \log \frac {a_{ji}  a_{j-1\,i-1}}{a_{j-1\, i} a_{j\, i-1}} 
\ee
to better conform with our notation, note that using reality we can write
$$
(i-1\, i\, j-1\, j) =a_{j-1 \, i}a_{j\, i-1}-a_{j-1\, i-1}a_{ji}
$$
So \eqref{MHV-symb} reduces to the differential equation
\begin{eqnarray*}
d R_n &=&\sum_{ij}  \log (a_{j-1 \, i}a_{j\, i-1}-a_{j-1\, i-1}a_{ji})\, \rd \log \frac {a_{ji}  a_{j-1\,i-1}}{a_{j-1\, i} a_{j\, i-1}}\\
&=& \sum_{ij} \left( \log \left( 1-\frac{a_{j-1\, i-1}a_{ji}}{a_{j-1 \, i}a_{j\, i-1}}\right) +\log(a_{j-1 \, i}a_{j\, i-1})\right) \rd \log \frac {a_{ji}  a_{j-1\,i-1}}{a_{j-1\, i} a_{j\, i-1}}\\
&=& \sum_{ij} \rd \Li_2 \left(\frac {a_{ji}  a_{j-1\,i-1}}{a_{j-1\, i} a_{j\, i-1}}\right) -\frac12 \rd  (\log(a_{j-1 \, i}a_{j\, i-1}))^2 
+ \log(a_{j-1 \, i}a_{j\, i-1})\rd \log ({a_{ji}  a_{j-1\,i-1}})\\
&=&\rd\left(\sum_{ij}  \Li_2 \left(\frac {a_{ji}  a_{j-1\,i-1}}{a_{j-1\, i} a_{j\, i-1}}\right) +\frac12  \log(a_{j-1 \, i}a_{j\, i-1})
\log \left(\frac{a_{ji}  a_{j-1\,i-1}}{a_{j-1 \, i}a_{j\, i-1}
}\right)\right)
\end{eqnarray*}
using $d \Li_2 x= -\log (1-x) \, \rd \log x$ and some resumming.
This is not quite a sum of functions of
 of the invariant $u_{ji}=\frac {a_{ji}  a_{j-1\,i-1}}{a_{j-1\, i} a_{j\, i-1}}$ including the $\log^2$ term,  so this superficially depends on the scalings, but  becomes independent of the scaling of the individual twistors only in the sum.

\end{document}